\documentclass[useAMS,usenatbib,usegraphicx,twocolumn]{mn2e}
\usepackage{color}
\usepackage{multirow}

\def\be{\begin{equation}}
\def\ee{\end{equation}}

\def\bh{M_{\bullet}}
\def\mbh{M_{\bullet}}
\def\mbht{M_{\bullet,\rm t}}
\def\mbhvir{M_{\bullet,\rm vir}}
\def\msun{M_{\odot}}
\def\msunyr{M_{\odot} {\rm yr^{-1}}}
\def\ergps{\rm erg~s^{-1}}
\def\kms{\rm km~s^{-1}}
\def\bol{_{\rm bol}}
\def\acc{_{\rm acc}}
\def\bol{_{\rm bol}}
\def\Mdot{\dot{M}_{\bullet,\rm acc}}

\title[Radiative Efficiency of SDSS QSOs]{Radiative Efficiency of Disk
Accretion in Individual SDSS QSOs}

\author[S. Wu et al.]{Shumei Wu$^{1,2}$, Youjun Lu$^1$\thanks{To whom
correspondence should be addressed. Email: luyj@nao.cas.cn}, 
Fupeng Zhang$^{1,2}$, \& Ye Lu$^1$ \\ 
$^{1}$National Astronomical Observatories, Chinese Academy of Sciences, 
Beijing, 100012, China\\
$^{2}$Graduate School of the Chinese Academy of Sciences, Beijing 100049, 
China }

\begin{document}

\maketitle

\begin{abstract}

We estimate the radiative efficiency $\epsilon$ of individual type 1
SDSS QSOs by using their bolometric luminosities ($L\bol$) and
accretion rates ($\Mdot$), which may be related to the assembly
histories and spins of the central massive black holes (MBHs). We
estimate $L\bol$ by using the empirical spectral energy distributions
of QSOs and $\Mdot$ by fitting the observed optical luminosity(/-ies)
with the thin accretion disk model, assuming the MBH masses given by
the virial mass estimator(s) ($\mbhvir$). We find an apparent
correlation between $\epsilon$ and $\mbhvir$, which is strong at
redshift $z\la1.8$, weak at $z\ga2$, and consistent with that found by
\citet{Davis11} for $80$ PG QSOs at $z\leq0.5$. To investigate whether
this correlation is intrinsic or not, we construct a mock sample of
QSOs according to the true MBH mass and Eddington ratio distributions
given in \citet{Kelly13}. By comparing the results obtained from the
mock sample with that from the SDSS sample, we demonstrate that the
apparent $\epsilon-\mbhvir$ correlation can be produced by and mainly
due to the selection effects of the SDSS sample and the bias induced
by the usage of $\mbhvir$ as the true MBH mass.  The mean values of
$\epsilon$ of those SDSS QSOs are consistent with being a constant
$\simeq0.11-0.16$ over the redshift range of $0.3\la z\la4$. We conclude 
that the current SDSS
QSO data is consistent with no strong intrinsic correlation between
radiative efficiency and true MBH mass and no significant redshift
evolution of radiative efficiencies.

\end{abstract}


\section{Introduction}

Disk accretion of gaseous material onto a massive black hole (MBH) is
believed to be the central engine of QSO and Active Galactic Nucleus
\citep[AGN\footnote{Hereafter, we do not distinguish AGN from QSO.};
e.g.,][]{Krolik99}. In the thin disk accretion scenario, the emergent
spectrum of a QSO is mainly determined by several parameters,
including the MBH mass ($\bh$) and spin ($a$), the accretion rate
($\dot{M}\acc$), and the inclination angle ($i$) of the disk to the
line of sight \citep[e.g.,][see also \citealt{Shakura73,NT73}
]{Krolik99}. Measuring the MBH mass ($M_{\bullet}$) and spin ($a$) for
each individual QSOs is of fundamental importance and has been one of
the major goals for the QSO and MBH studies. As illustrated by a
number of studies \citep[e.g.,][]{Gammie04, Volonteri05, King06,
King08, Berti08, Volonteri12}, obtaining the statistical distributions of
$M_{\bullet}$ and $a$ among QSOs is probably crucial for our
understanding of the growth history of MBHs and the cosmic evolution
of QSOs. In principle, $\bh$, $\dot{M}\acc$ and $a$ can be estimated
through the fitting of the multi-wavelength spectrum of each QSO by
dedicated disk model(s) \citep[e.g.,][]{SM89}.  However, it is
practically not easy to estimate these parameters simultaneously with
considerable accuracy, partly because of the complications in disk
accretion model(s) and partly because of lack of measurements at the
EUV to soft X-ray bands, where the disk emission is expected to be
most prominent, for most QSOs.  Alternative methods to estimate $\bh$
and $a$ are necessary for understanding the evolution and assembly
history of MBHs.

The mass of a MBH in a QSO can be estimated through the reverberation
mapping (RM) technique \citep[][]{BM82}, which is time consuming and
currently is not possible for all QSOs. Fortunately, recent
developments in measuring the masses of MBHs through the RM technique
have resulted in some empirical scaling relations, i.e., the virial
mass estimators, which offer simple ways to estimate the MBH mass of a
QSO \citep[e.g.,][]{Kaspi00, Peterson04, Vestergaard02, Vestergaard06,
Shen08, Shen11}. Adopting these virial mass estimators, \citet{Shen08,
Shen11} have obtained the MBH masses for most SDSS QSOs by using the
QSO optical luminosities and the width of some broad emission lines
(such as H$\beta$, CIV, and Mg II) in the QSO spectra. 

If the MBH mass is known for a QSO, the absolute accretion rate
$\dot{M}\acc$ can be estimated based on the optical band luminosity by
using the thin disk accretion model. The radiative efficiency is then
obtained by $\epsilon =L\bol/\dot{M}\acc c^2$, where $L\bol$ is the
bolometric luminosity. The MBH spin $a$ can be inferred from
$\epsilon$ as $\epsilon$ is simply determined by $a$ in the thin disk
accretion model.  With this method, it may be possible to indirectly
infer the MBH spin and its distribution for a large sample of QSOs.
(Note here that direct measuring the spins of MBHs through features
like the broad Fe K$\alpha$ line is currently possible only for a
limited number of cases (e.g., MCG-6-30-15 and a dozen of other AGNs;
see \citealt{Fabian00,Risaliti13,Reynolds13a,Reynolds13b}; and
reference therein).

\citet{Davis11} estimated $\epsilon$ for $80$ PG QSOs (hereafter DL
sample or DL QSOs) by adopting the above method and found that
$\epsilon$ is strongly correlated with $\bh$, i.e., $\epsilon \propto
\bh^{0.5}$, and they argued that such a correlation is unlikely
induced by selection effects \citep[][see also
\citealt{Chelouche13}]{LD11}. If the above correlation is intrinsic,
it might suggest that the spin of a big MBH is relatively larger than
that of a small MBH and the growth histories of big MBHs are
systematically different from that of small MBHs. However,
\citet{Raimundo12} argued that such a correlation is most likely an
artifact of the small parameter space in both luminosity and redshift
covered by the DL sample, and additional factors, such as the host
galaxy and dust contamination, uncertainties in the bolometric
luminosity estimations, may also bias the estimations of $\epsilon$.
\citet{Raimundo12} concluded that the radiative efficiency itself
cannot be accurately measured due to the large errors in the relevant
observed parameters.  Therefore, it is still not clear whether
$\epsilon$ really intrinsically correlates with $\bh$ or not. 

In this study, we extend the work of \citet{Davis11} to the type
1 SDSS QSOs, in order to further study the radiative efficiency of
individual QSOs and its relation, if any, with MBH mass. Such a study
should be helpful to identify various biases involved in the
estimations of $\epsilon$ and address the question of whether
$\epsilon$ correlates with $\bh$ or not, as the SDSS sample has
many more QSOs and covers larger luminosity and redshift ranges
compared to the DL sample.

This paper is organized as follows. In section~\ref{sec:SDSSQSO}, we
introduce the SDSS QSO sample adopted in this study. In
section~\ref{sec:estepsilon}, we first introduce the method to
estimate the radiative efficiency and then obtain the radiative
efficiency for individual type 1 SDSS QSOs. Similar to
\citet{Davis11}, we also find that the estimated radiative efficiency
appears to be strongly correlated with the MBH virial mass at redshift
$z\leq 0.5$. Such a strong correlation appears to exist for the SDSS
QSOs at $z< 1.9$ but not at $z\ga 2$. In section~\ref{sec:bias}, we
investigate various biases that could be involved in the $\epsilon$
estimations and the selection effects of the QSO sample, and adopt
Monte Carlo simulations to generate mock QSO samples to simulate the
effects of these biases. We find that the current data is consistent
with no mass and redshift dependence of $\epsilon$. Conclusions are
given in section~\ref{sec:conclusions}.

\section{SDSS QSO sample}
\label{sec:SDSSQSO}

A large number of QSOs have been detected by the SDSS. In the catalog
of SDSS QSOs (DR7), $104,746$ QSOs have $i$-band magnitude $M_i< -22$,
at least one broad emission line broader than $1000\kms$, and the
estimations of the central MBH masses. About half of these QSOs (totally
$57,959$ QSO at $0.3\leq z\leq 5$) were selected out to form a
homogeneous, statistical sample, which is primarily a flux limited
sample with $i$-band magnitude $m_i\leq 19.1$ at $z\leq 2.9$, $m_i
\leq 20.2$ at $z>2.9$ and an additional bright limit of $m_i\geq 15$
\citep{Richards02,Kelly13}. Estimations of the MBH masses of the QSOs
in this sample have been obtained by using the virial mass estimators
based on scaling relationships calibrated for four different emission
lines, i.e., H$\alpha$, H$\beta$, Mg II, and CIV. These estimations,
together with the optical band luminosities, for this QSO sample are
given in \citet{Shen11}. We select the QSOs with $0.3\leq z \leq 4$ in
this sample to form our QSO sample (totally $56,691$ QSOs) to be
studied in this work, and these QSOs are denoted as the SDSS QSOs hereafter if
not otherwise stated. With this homogeneous, statistical QSO sample,
it is possible to study and model the selection effects and various
biases involved in the estimations of the radiative efficiency of
individual QSOs as the true MBH mass function and Eddington ratio
distribution have also been estimated by \citet[][see details in
section~\ref{sec:bias}]{Kelly13}. 

The monochromatic luminosities are provided only at one or two
wavelengths for most of the QSOs in the sample of \citet{Shen11}. For
example, the monochromatic luminosities at $5100{\rm \AA}$, $3000{\rm
\AA}$, and $1350{\rm \AA}$ are provided for QSOs with redshift
$z<0.9$, $0.35<z<2.25$, and $z>1.5$, respectively. For those QSOs with
$0.35<z<0.9$, there are two monochromatic luminosity
measurements available at $5100{\rm \AA}$ and $3000{\rm \AA}$,
respectively; and for those QSOs with $1.5<z<2.25$, there are
also two monochromatic luminosity measurements available at $3000{\rm
\AA}$ and $1350{\rm \AA}$, respectively.  For other QSOs, only one
monochromatic luminosity measurement is available, either at
$5100{\rm \AA}$ for those with $z<0.35$ or at $1350{\rm \AA}$ for those
with $z>2.25$.

\section{Radiative efficiencies of disk accretion in individual SDSS
QSOs} \label{sec:estepsilon}

The radiative efficiency ($\epsilon$) of the accretion process in a
QSO may be estimated by using its bolometric luminosity ($L_{\rm
bol}$) and the accretion rate ($\dot{M}\acc$), i.e., $\epsilon =
L\bol/\dot{M}\acc c^2$, if these two quantities can be
(independently) accurately determined. Given the mass of the central
MBH of a QSO and the QSO multi-wavelength spectrum, in principle, the
accretion rate can be estimated by adopting dedicated disk accretion
model(s); and the bolometric luminosity can also be estimated by
integrating the QSO multi-wavelength spectrum. In this section, we will
first estimate the mass accretion rate and the bolometric luminosity
for individual SDSS QSOs, and then estimate the radiative efficiency
$\epsilon$.

\subsection{Mass accretion rates $\dot{M}\acc$}
\label{subsec:mdot}

In the standard disk accretion scenario, the multi-wavelength spectrum
of a QSO is mainly determined by the mass and spin of the central MBH
and the accretion rate \citep[e.g.,][see also \citealt{Shakura73,NT73}
]{Krolik99}. The masses of the central MBHs in most SDSS QSOs have
been estimated through the virial estimators \citep[e.g.,][]{Shen08,
Shen11}, while the MBH spins are still difficult to measure.  The
optical band luminosity of a QSO is dominated by the emission from the
outer disk, and it is mainly determined by the accretion rate of the
disk and less affected by the MBH spin \citep[e.g.,][]{Krolik99,
Davis11}. Therefore, the accretion rate $\dot{M}\acc$ of a SDSS QSO,
with known MBH mass, can be estimated with reasonable accuracy by only
matching the observed optical band luminosity(/-ies) $L_{\rm opt}$
with that predicted by an accretion disk model.  Similar to
\citet{Davis11}, we adopt two different accretion disk models, i.e.,
the standard thin disk model and the TLUSTY model to estimate the mass
accretion rate of individual SDSS QSOs as follows.

\begin{itemize}

\item The standard thin accretion disk model (hereafter the BB model).
Here we adopt the relativistic standard thin disk model presented in
\citet[][see also \citealt{NT73}, \citealt{Page74}]{Gierlinski01}. In
this model, the inner edge of the accretion disk is assumed to be the
innermost stable circular orbit (ISCO), which is solely determined by
the MBH spin, and no emission comes from the region within ISCO. The
emergent spectrum is directly determined by the multi-temperature
black body radiation integrated over the disk surface.

\item The TLUSTY model \citep{Hubeny00,HL95}. In this model, not only
the relativistic effects are included, but also the disk vertical
structure and the radiative transfer in the disk are simultaneously
considered \citep[for details, see][]{Hubeny00}. In the TLUSTY model,
the emergent spectrum from each annulus is first calculated and the
full spectrum is obtained by integrating over the disk radius.
Occasionally, the TLUSTY model could not converge for some specific
parameter space, e.g., at large radius or extremely large/small
accretion rate, possibly because the disk material in this situation
is already quite cool and/or convection becomes important. In these
cases, we simply replace the radiation of these annuli by black body
radiation \citep[see also discussions in][]{Davis11}.

\end{itemize}

These models have four parameters, i.e., the MBH mass $\mbh$ and spin
$a$, the accretion rate $\dot{M}\acc$, and the inclination angle $i$
of the disk to the line of sight. The model spectrum estimated from
the TLUSTY model may also depend on the choice of the $\alpha_{\rm
SS}$-viscosity \citep{Shakura73}, which is simply set to be
$\alpha_{\rm SS}=0.01$ for all models as this parameter have little
effect on the $L_{\rm opt}$ \citep[e.g., see][]{Davis11}.

It is necessary to have some information about the other model
parameters, i.e., $\mbh$, $a$ and $i$, in order to estimate
$\dot{M}\acc$ with reasonable accuracy for a QSO by only using its
monochromatic luminosities in a limited range of wavelengths.  The MBH
mass, $\mbh$, has been estimated for the majority of the SDSS QSOs by
using the virial mass estimators but with uncertainties of
$0.3-0.4$~dex \citep[e.g.,][]{Shen08, Shen11}.  The inner radius of an
accretion disk is defined by the MBH spin $a$.  The model spectra in
the optical bands are not significantly affected by the setting of the
disk inner edge although the disk radiation at higher frequency is
indeed affected by it. If not otherwise stated, we simply adopt
$a=0.67$ in our following calculations, which corresponds to
$\epsilon\sim 0.1$ in the standard disk model.\footnote{Alternatively
adopting $a=0.83$ (or $a=0.96$), i.e., $\epsilon \sim 0.13$ (or $0.2$), does not significantly
affect the final results.}  We adopt this spin value as reference
since the global constraints on the mean radiative efficiency of QSOs
suggest $\epsilon \sim 0.1-0.2$ \citep[e.g.,][]{YT02, YL04, Marconi04,
YL08, Shankar09, Zhang12, Shankar13}.  For the inclination angle $i$,
all QSOs in the sample of \citet{Shen11} are type 1 QSOs, which
presumably have small inclination angles with $\cos i$ in the range of
$0.5$ to $1$. The exact value of $i$ for each QSO cannot be
constrained directly from the SDSS observations. In this section, we
set a fixed value of $\cos i = 0.8$ for the fitting, with which the
results obtained here can be directly compared with that found in
\citet{Davis11} and \citet{Raimundo12}.

With the virial mass of a central MBH and the above settings on the
MBH spin and inclination angle, it is possible to estimate the
accretion rate for the QSO by matching the observed monochromatic
luminosity in an optical band to the model ones. In the sample of
\citet{Shen11}, the monochromatic luminosities are provided at two
wavelengths for those QSOs with $0.35<z<0.9$ or $1.5<z<2.25$ (i.e., at
$5100$\AA~ and $3000$\AA, or at $3000$\AA~ and $1350$\AA;
see section~\ref{sec:SDSSQSO}). For these cases, the two estimates of
the accretion rate generally differ by no more than $0.1-0.2$~dex
according to our calculations. We adopt the mean value of the two
estimations of the accretion rate, obtained from the two monochromatic
luminosity measurements, as the QSO accretion rate, and the
uncertainty of this rate is $\la 0.1$~dex, which may lead to an
uncertainty of $\la 0.1$~dex in the efficiency estimation. However,
such an amount of uncertainty seem negligible compared with the
biases induced by the usage of the virial mass as the true MBH mass
as mainly discussed in section~\ref{subsec:bias_m} below.

The obtained value of $\dot{M}\acc$ for a QSO may be biased because of
the usage of the virial mass as the true MBH mass and the fixed values
for the MBH spin and the disk inclination angle here. The biases
induced by these settings on $\dot{M}\acc$ and consequently the
$\epsilon$ estimations will be further analyzed in
section~\ref{sec:bias}.

\subsection{Bolometric luminosities $L_{\rm{bol}}$}
\label{subsec:lbol}

It is necessary to estimate its bolometric luminosity with
considerable accuracy to obtain the radiative efficiency $\epsilon$ of
a QSO.  For a QSO with multi-wavelength observations, its bolometric
luminosity can be directly obtained by integrating its spectrum over
the range from radio to hard X-ray but excluding the reprocessed
radiation in the infrared. However, most SDSS QSOs have been observed
only in a limited wavelength range, e.g., in the optical band. It may
not be an appropriate way to directly convert the optical band
luminosity to the bolometric luminosity according to a theoretical
accretion disk model, as the radiation processes for high energy
photons (which contribute significantly to the total luminosity)
emitted from the inner disk region are complicated and not fully
understood \citep[e.g.,][]{Krolik99}. Here we adopt an empirical
method to calculate the bolometric luminosity of individual QSOs by
constructing a spectral energy distribution (SED) for individual QSOs
based on some empirical relations obtained for small samples of QSOs,
which is similar to that adopted in \citet{Hopkins07} and \citet[][and
see references therein]{Marconi04}. 

\begin{figure}
\centering
\includegraphics[scale=0.85]{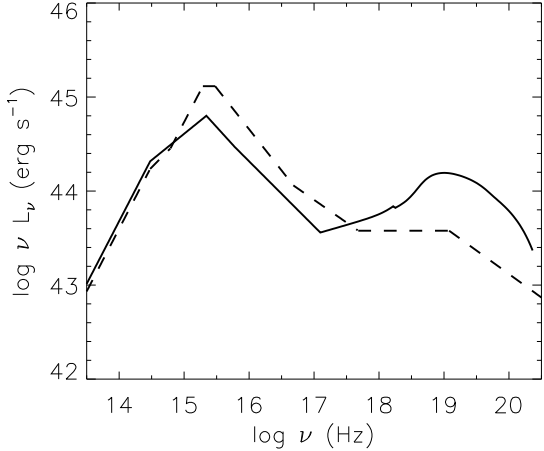}

\caption{Schematic spectral energy density distribution (SED) adopted
for a QSO with $L_{2500 \rm \scriptsize \AA}=10^{44.5}\ergps$. The
solid and dashed lines show an example SED of that adopted in this
study (see section~\ref{subsec:lbol}) and that adopted in \citet[][see case A in
section 3.1 therein]{Davis11}, respectively.  }

\label{fig:f1}
\end{figure}

\begin{figure}
\centering
\includegraphics[scale=0.85]{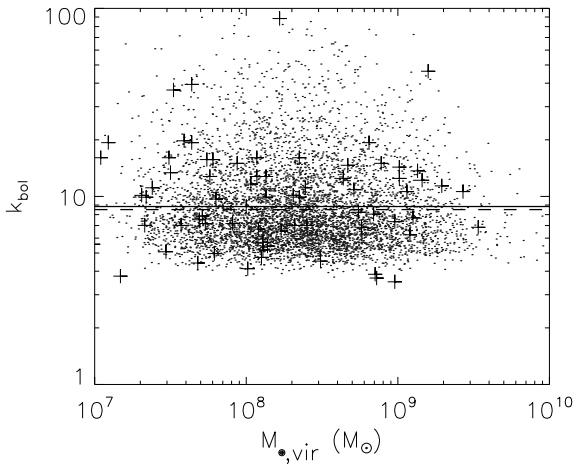}

\caption{Bolometric corrections $k\bol$ at $5100$\AA\ for the type 1
SDSS QSOs at redshift $z\leq 0.5$ (black dots; totally 5,786 SDSS
QSOs) and the DL QSOs (black crosses; adopted from \citealt{Davis11}),
respectively. Solid and dashed lines represent the mean value of $\log
k\bol$ for the SDSS QSOs ($\left<\log k\bol\right> =0.95$) and the DL
QSOs ($\left<\log k\bol \right> =0.93$), respectively.  }

\label{fig:f2}                                                
\end{figure}                                                  

The spectra or SEDs of individual QSOs are constructed as follows. 

\begin{enumerate}

\item In the optical-UV bands ( $1\mu$m$<\lambda<1300$\AA), the QSO
spectrum is assumed to be a power law, and we assign the power-law
index to each QSO according to a Gaussian distribution with mean
$\alpha_{\rm opt}=-0.44$ with scatter $\sigma_{\alpha_{ \rm
opt}}=0.125$ \citep[see][]{VandenBerk}.  The spectrum is normalized to
the monochromatic luminosity at $5100$\AA, i.e., $L_{5100\rm
\scriptsize \AA}$. (If $L_{5100\scriptsize \rm \AA}$ is not available
for some QSOs, we estimate it by extrapolation from $L_{3000\rm
\scriptsize \AA}$ or $L_{1350\rm \scriptsize \AA}$).  

\item In the UV wavelengths ($1200$\AA-$500$\AA), the QSO spectrum is
also described by a power law, and we randomly assign the power-law
index according to a Gaussian distribution with mean $\alpha_{\rm
UV}=-1.76$ with a scatter of $\sigma_{\alpha_{\rm UV}}=0.12$
\citep{Telfer02}.  

\item At the X-ray beyond $0.5$keV, the spectrum is assumed to be a
power law with an exponential cutoff at $500$keV. We assign the photon
index according to a Gaussian distribution with mean $\Gamma=-1.83$
and a scatter of $\sigma_{\Gamma}=0.18$ for each QSO \cite[][see also
\citealt{Tozzi06}]{JinDone12}.

\item  A reflection component is generated for each QSO by using the
PEXRAV model \citep{Magdziarz95} in the XSPEC package with a
reflection solid angle of $2\pi$, inclination of $\cos i=0.8$ and
solar abundance. 

\item The X-ray spectrum is re-normalized to a given $\alpha_{\rm
ox}=-0.384 \log \left[L_{2500 \rm \scriptsize \AA}/L_\nu(2{\rm
keV})\right]$, and the points at $500$\AA \ and $50$\AA \ are
connected with a power-law.  The value of $\alpha_{\rm ox}$ depends on
luminosity, and we adopt the most recent determination by \citet[][see
also \citealt{Steffen06}]{Lusso10}, i.e., $
\alpha_{\rm{ox}}=-0.154\log (L_{2500\rm \scriptsize \AA} /{\rm erg
s^{-1} Hz^{-1}})+3.176,$ with an intrinsic scatter of
$\sigma_{\alpha_{\rm ox}}=0.18$. 

\item At wavelengths longer than $1\mu$m, the QSO spectrum may be
dominated by the emission from the dusty torus by re-processing the
optical-UV-X-ray photons radiated from the inner disk, which should
not be counted in the calculations of the bolometric luminosities of
QSOs. We assume a power law continuum emission with slope $1/3$,
similar to that adopted in \citet{Davis11}.  

\end{enumerate}

\begin{figure*}
\centering
\includegraphics[scale=0.8]{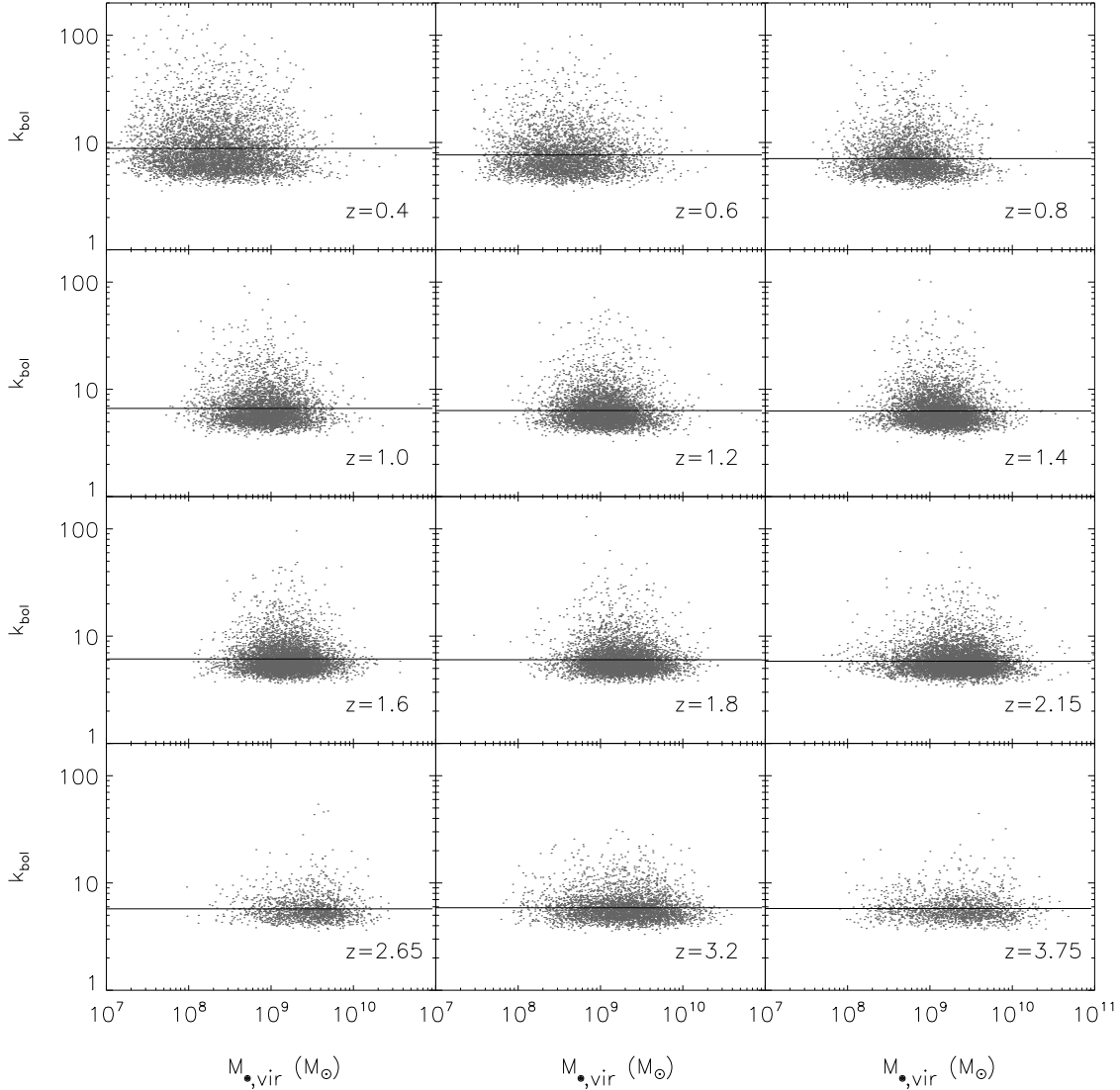}

\caption{Bolometric corrections $k\bol$ at $5100$\AA\ for the type 1
SDSS QSOs in different redshift bins. The solid line represents the
mean value of $\left<\log k\bol\right>$ for the SDSS QSOs in each
redshift bin. For those QSOs with monochromatic luminosities available
at $3000$\AA\ and/or $1350$\AA, we obtain $L_{5100\scriptsize \rm \AA}$
according to the the empirical spectral energy distributions.  }

\label{fig:f3}
\end{figure*}

Figure~\ref{fig:f1} shows an example SED constructed for a QSO with
$L_{2500 \rm \scriptsize \AA}=10^{44.5}\ergps$ according to the above
prescription. As seen from Figure~\ref{fig:f1}, the general SED
adopted in this study (the solid line) is somewhat different from that
adopted in \citet[][the dashed line]{Davis11}, and the differences
include (1) the reflection component of the X-ray emission due to the
disk is considered in this study but not in \citet{Davis11}; (2) our
choice of the intrinsic X-ray continuum above $0.2$~kev is also
slightly different from that in \citet{Davis11}; (3) the continuum
slope at optical to UV band is set to be $-0.44$ in our study; but in
\citet{Davis11} it is set to be $-0.3$ between $1\mu$m and $4861$\AA,
$-1$ between $1549$\AA~ and $1000$\AA~, and set to be the one measured
for each QSO in the wavelength range from $4861$\AA~ to $1549$\AA~ and
from $1000$\AA~ to $0.2$~kev, respectively.  However, these
differences do not lead to much difference in the bolometric
corrections as shown in Figures~\ref{fig:f2} and \ref{fig:f3}.

By integrating over the constructed spectrum of each QSO in the sample
according to the above prescription, we obtain its ``bolometric
luminosity'', and the distribution of the ``bolometric luminosity'' of
these QSOs should statistically reflect the underlying real
distribution of the sample. Given the bolometric luminosity, we also
obtain the bolometric correction (BC) at $5100$\AA\ for each QSO with
the luminosity measurement at $5100$\AA, i.e., $k_{\rm bol} = L\bol
/L_{5100\rm \scriptsize \AA}$.

Figure~\ref{fig:f2} shows the BCs at $5100$\AA\ obtained for the SDSS
QSOs at redshift $z\leq 0.5$. As seen from Figure~\ref{fig:f2}, the
obtained BCs do not depend on $\mbhvir $, which is consistent with the
results obtained by \citet{Davis11} for a sub-sample of PG QSOs in a
similar redshift range (the plus symbols in Figure~\ref{fig:f2}). The
range of the BCs for the SDSS QSOs is also roughly consistent with
that for the DL sample. The mean logarithmic BC value of the DL QSOs
is $\left<\log k\bol\right>=0.93$ with a scatter of $\sigma_{\log
k\bol} \simeq 0.27$, while that of the SDSS QSOs is $\left<\log k\bol
\right> =0.95$ with a scatter of $\sigma_{\log k\bol} \simeq 0.22$ (as
shown by the solid and dashed lines in Figure~\ref{fig:f2},
respectively). The mean value of the logarithmic BCs for the SDSS QSOs
is slightly larger than that of the DL sample, which might be due to the
choice of the empirical SED.  For example, if we choose $\alpha_{\rm
ox}=-0.137\log (L_{2500\rm \scriptsize \AA} /{\rm erg s^{-1}
Hz^{-1}})+2.638 $ as given by \citet{Steffen06}, then we would have
$\left<\log k\bol\right>=0.88$ for the SDSS QSOs. Other estimations of
the BCs for QSOs at $5100$\AA\ include $k\bol \sim 9$
\citep[][]{Kaspi00}, $k\bol \sim 10.3\pm 2.1$ \citep[][]{Richards06b},
and $k\bol \sim 8.1\pm 0.4$ \citep{Runnoe12}, etc., which are roughly
consistent with that shown in Figure~\ref{fig:f2} and suggest that the
systematic uncertainties in the BCs should not exceed $\pm 0.06$~dex.

\begin{table*}
\begin{center}
\begin{tabular}{cccccccccccc} \hline
\multirow{2}{*}{z} & \multicolumn{4}{c}{The BB model} &  & \multicolumn{4}{c}{The TLUSTY model} & 
\multirow{2}{*}{$N_{\rm Q}$} \\ \cline{2-5} \cline{7-10} 
  & $R_{\rm S}$ & $\alpha$ & $\beta$ & $\left<\log\epsilon \right>$ &  &
$ R_{\rm S}$ & $\alpha$ & $\beta$ & $\left<\log\epsilon \right>$ & \\ \hline
0.3-0.5 &   0.79 & -1.15 &   0.64 & -0.91$\pm$0.41 &  &  0.72 & -1.15 & 0.56 & -0.95$\pm$0.38 & 4,291 \\
0.5-0.7 &   0.71 & -1.25 &   0.55 & -0.93$\pm$0.35 &  &  0.66 & -1.24 & 0.50 & -0.95$\pm$0.34 & 4,205 \\
0.7-0.9 &   0.62 & -1.30 &   0.47 & -0.95$\pm$0.30 &  &  0.60 & -1.32 & 0.45 & -0.98$\pm$0.30 & 3,953 \\
0.9-1.1 &   0.69 & -1.24 &   0.45 & -0.82$\pm$0.24 &  &  0.70 & -1.28 & 0.45 & -0.87$\pm$0.24 & 4,861 \\
1.1-1.3 &   0.70 & -1.26 &   0.43 & -0.82$\pm$0.21 &  &  0.70 & -1.31 & 0.44 & -0.85$\pm$0.23 & 5,865 \\
1.3-1.5 &   0.69 & -1.29 &   0.43 & -0.81$\pm$0.20 &  &  0.70 & -1.36 & 0.45 & -0.85$\pm$0.22 & 5,903 \\
1.5-1.7 &   0.66 & -1.26 &   0.37 & -0.81$\pm$0.19 &  &  0.71 & -1.36 & 0.43 & -0.85$\pm$0.21 & 6,451 \\
1.7-1.9 &   0.53 & -1.14 &   0.26 & -0.81$\pm$0.17 &  &  0.64 & -1.28 & 0.35 & -0.86$\pm$0.20 & 5,840 \\
1.9-2.4 &   0.22 & -1.04 &   0.15 & -0.85$\pm$0.18 &  &  0.37 & -1.15 & 0.22 & -0.87$\pm$0.20 & 7,734 \\
2.4-2.9 &   0.07 & -0.97 &   0.06 & -0.87$\pm$0.13 &  &  0.25 & -1.08 & 0.13 & -0.89$\pm$0.14 & 1,656 \\
2.9-3.5 &   0.17 & -0.99 &   0.08 & -0.88$\pm$0.16 &  &  0.34 & -1.07 & 0.14 & -0.89$\pm$0.18 & 4,176 \\
3.5-4.0 &   0.17 & -1.11 &   0.08 & -0.90$\pm$0.15 &  &  0.32 & -1.07 & 0.14 & -0.90$\pm$0.18 & 1,756 \\ \hline
\end{tabular}
\end{center}
\caption{Efficiency versus MBH virial mass relation for the SDSS QSOs
in different redshift bins. The relationship between $\epsilon$ and
$\mbhvir $, if any, is fitted by $\log\epsilon=\alpha+\beta\log
(\mbhvir/10^8\msun)$, $\alpha$ and $\beta$ are the two parameters of
the linear fit. The first column is for the redshift range of the
QSOs.  Symbols $R_{\rm S}$ and $\left<\log \epsilon\right>$  represent
the Spearman rank correlation coefficient and the mean value of the
logarithmic radiative efficiency for the SDSS QSOs in each redshift
bin, and $N_{\rm Q}$ represents the total number of the QSOs in each
redshift bin. For the SDSS QSOs in each redshift bin, $R_{\rm S}$,
$\alpha$, $\beta$, and $\left<\log \epsilon\right>$ are obtained by
adopting either the BB model or the TLUSTY model as listed in the
second to fifth columns and the sixth to ninth columns,
respectively.} 
\label{tab:t1}
\end{table*}

Figure~\ref{fig:f3} shows the BCs at $5100$\AA\ obtained for the SDSS
QSOs in different redshift bins, and these redshift bins are separated
by boundaries of $0.3$, $0.5$, $0.7$, $0.9$, $1.1$, $1.3$, $1.5$,
$1.7$, $1.9$, $2.4$, $2.9$, $3.5$ and $4.0$ \citep[see][]{Kelly13}.
As seen from Figure~\ref{fig:f3}, the BCs do not depend on $\mbhvir$
in all the redshift bins; and the BCs in different redshift bins
spread in roughly the same range, though the mean value of the BCs
decreases slightly from $\sim 9$ to $\sim 6$ with redshift increasing
from $z\sim 0.4$ to $z\sim 3.75$. The increasing of the mean value of
BCs with redshift is primarily a consequence of that $\alpha_{\rm ox}$
decreases with increasing luminosity and the high redshift QSOs in
general have larger luminosities.

\subsection{Radiative efficiencies of SDSS QSOs}

\begin{figure*}
\centering
\includegraphics[scale=0.8]{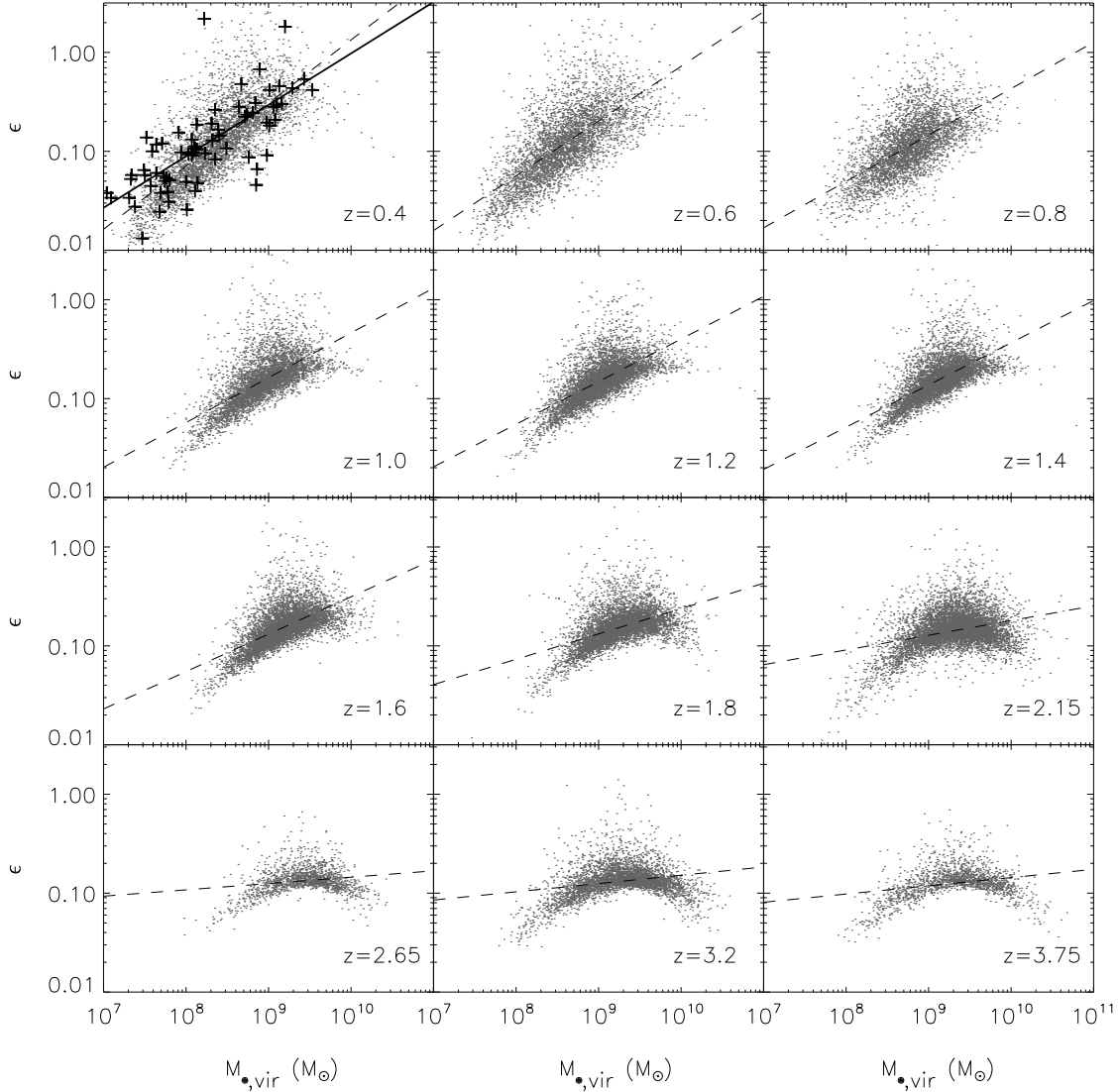}

\caption{Distributions of the SDSS QSOs in the $\epsilon$--$\mbhvir$
plane. The twelve panels show the results of the radiative efficiency
($\epsilon$) for the SDSS QSOs in different redshift bins, i.e., $0.3
\leq z<0.5$, $0.5\leq z<0.7$, $0.7\leq z<0.9$, $0.9\leq z<1.1$,
$1.1\leq z<1.3$, $1.3\leq z<1.5$, $1.5\leq z<1.7$, $1.7\leq z< 1.9$,
$1.9\leq z<2.4$, $2.4\leq z<2.9$, $2.9\leq z<3.5$, and $3.5\leq
z<4.0$, respectively. Each grey point represents an SDSS QSO, and each
plus symbol represents a QSO in the DL sample estimated by
\citet{Davis11}. The dashed line in each panel represents the best
linear fit of the relationship between $\log \epsilon$ and $\log
\mbhvir $ in each particular redshift bin.  The dependence of
$\epsilon$ on $\mbhvir $ is strong at redshift $z \la 1.8$ but becomes
weak at redshift higher than $2$. The solid line represents the
fitting for the DL QSOs \citep{Davis11}.  }

\label{fig:f4}
\end{figure*}

In this section, we estimate the radiative efficiency for each SDSS
QSO in our sample (see section~\ref{sec:SDSSQSO}).  We first estimate
the $\dot{M}\acc$ for each QSO by matching the observational
monochromatic luminosity at $5100$\AA\ (or $3000$\AA, or $1350$\AA)
with both the BB model and the TLUSTY model (see
section~\ref{subsec:mdot}). If monochromatic luminosities are
available at more than one wavelength, we obtain $\dot{M}\acc$ for
each monochromatic luminosity and take the mean value of the 
$\dot{M}\acc$ estimates as the accretion rate of the QSO. We also
estimate the bolometric luminosity $L\bol$ for each QSO by using the
empirical method described in section~\ref{subsec:lbol}. With the
estimated $\dot{M}\acc$ and $L\bol$, the radiative efficiency of a QSO
can be then obtained through $\epsilon=L\bol / \dot{M}\acc c^2$. 

Figure~\ref{fig:f4} shows the distribution of the SDSS QSOs in the
$\epsilon$--$\mbhvir $ plane, where $\epsilon$ is estimated through
the BB model. From left to right and top to bottom, each panel shows
the results for a given redshift bin, and these redshift bins are
separated by boundaries of $0.3$, $0.5$, $0.7$, $0.9$, $1.1$, $1.3$,
$1.5$, $1.7$, $1.9$, $2.4$, $2.9$, $3.5$ and $4.0$
\citep[see][]{Kelly13}. As seen from Figure~\ref{fig:f4}, the
radiative efficiencies of the SDSS QSOs appear to be strongly
correlated with the masses of the central MBHs in each redshift bin
with $z< 1.9$. We perform Spearman's rank order correlation analysis
for the QSOs in each redshift bin and the results are listed in
Table~\ref{tab:t1}. If adopting the TLUSTY model, we obtain similar
results, see Table~\ref{tab:t1}.  Apparently, the correlation between
$\epsilon$ and $\mbhvir $ is strong in the redshift bins with $z\la
1.8$, while it becomes weak at redshift $z\ga 2$. We adopt a power-law
form to fit such a correlation, i.e., 
\begin{equation}
\log \epsilon =\alpha +\beta \log 
\left( \frac{\mbhvir}{10^8M_{\odot}}\right),
\end{equation}
where $\alpha$ and $\beta$ are the two fitting parameters. The
parameters that best fit the estimated $\log\epsilon$ versus $\log
\mbhvir $ relation in each redshift bin are listed in
Table~\ref{tab:t1}. For the lowest redshift bin ($0.3<z<0.5$),
$\beta=0.64$ and $0.56$ if adopting the BB model and the TLUSTY model,
respectively, which is roughly consistent with that obtained in
\citet[][$\beta=0.52$]{Davis11} for the DL QSOs in a similar redshift
range. The slight difference might be due to the SED shape adopted
here is different from that in the DL sample and the selection
criteria for the DL QSO sample is different from that for the SDSS
sample.  In the first panel ($z=0.4$), the DL QSOs in \citet{Davis11}
are also plotted (plus symbols) and apparently the distribution of
these DL QSOs are also roughly consistent with the SDSS QSOs. The
fitting value of the power-law slope ($\beta$) decreases from $0.64$
to $\sim 0$ with increasing redshift from $z\sim 0.2$ to $z\sim 4$
(see Table~\ref{tab:t1} and Figure~\ref{fig:f4}).  For the whole SDSS
QSO sample, the best fit gives $\beta =0.24$, which is substantially
smaller than the best fit of $\beta$ ($=0.64$) for the lowest redshift
bin ($0.3<z<0.5$).
 
Note other studies, e.g., \citet{CL08}, \citet{Wang09}, \citet{Li12},
and \citet{Shankar13}, in addition to \citet{Davis11}, also introduced
the dependence of the radiative efficiency of QSOs on the MBH mass
(and/or redshift), but based on the global integral properties of QSOs
through the So{\l}tan argument \citep{Soltan82,SB92,YT02}.

\begin{figure*}
\centering
\includegraphics[scale=0.85]{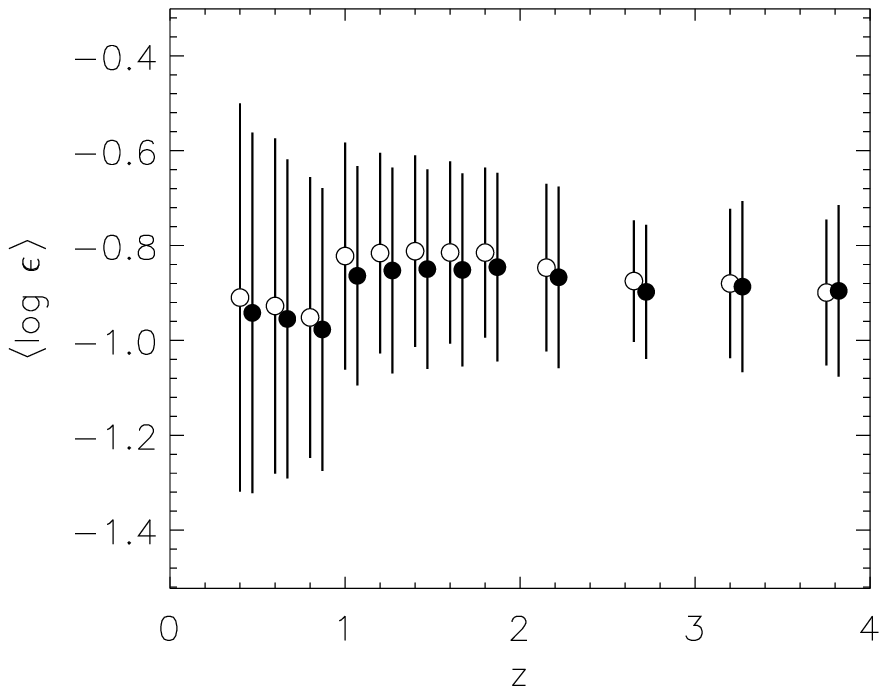}
\includegraphics[scale=0.85]{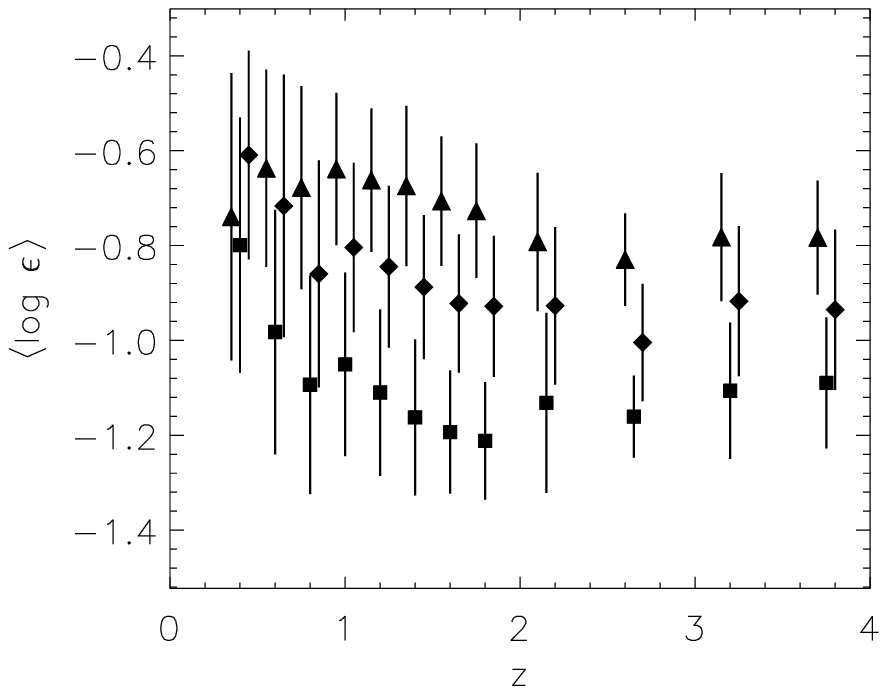}

\caption{Mean logarithmic radiative efficiency of the type 1 SDSS QSOs
in different redshift bins. The left panel shows the mean logarithmic
radiative efficiency of all the QSOs in each redshift bin obtained by
using either the BB model (open circles) or the TLUSTY model (solid
circles).  The right panel shows the mean logarithmic radiative
efficiency obtained by using the BB model for the QSOs in each
redshift bin with MBH masses in the range of $2.5\times       
10^8$--$4\times 10^8 \msun $ (squares), $8\times 10^8$--$1.3\times
10^9 M_{\odot}$ (diamonds) and $2.5\times 10^9$--$4 \times 10^9
M_{\odot}$ (triangles), respectively. The bars associated with each
symbol represent the standard deviation of the mean value.  }

\label{fig:f5}
\end{figure*}

\begin{figure*}
\centering
\includegraphics[scale=0.85]{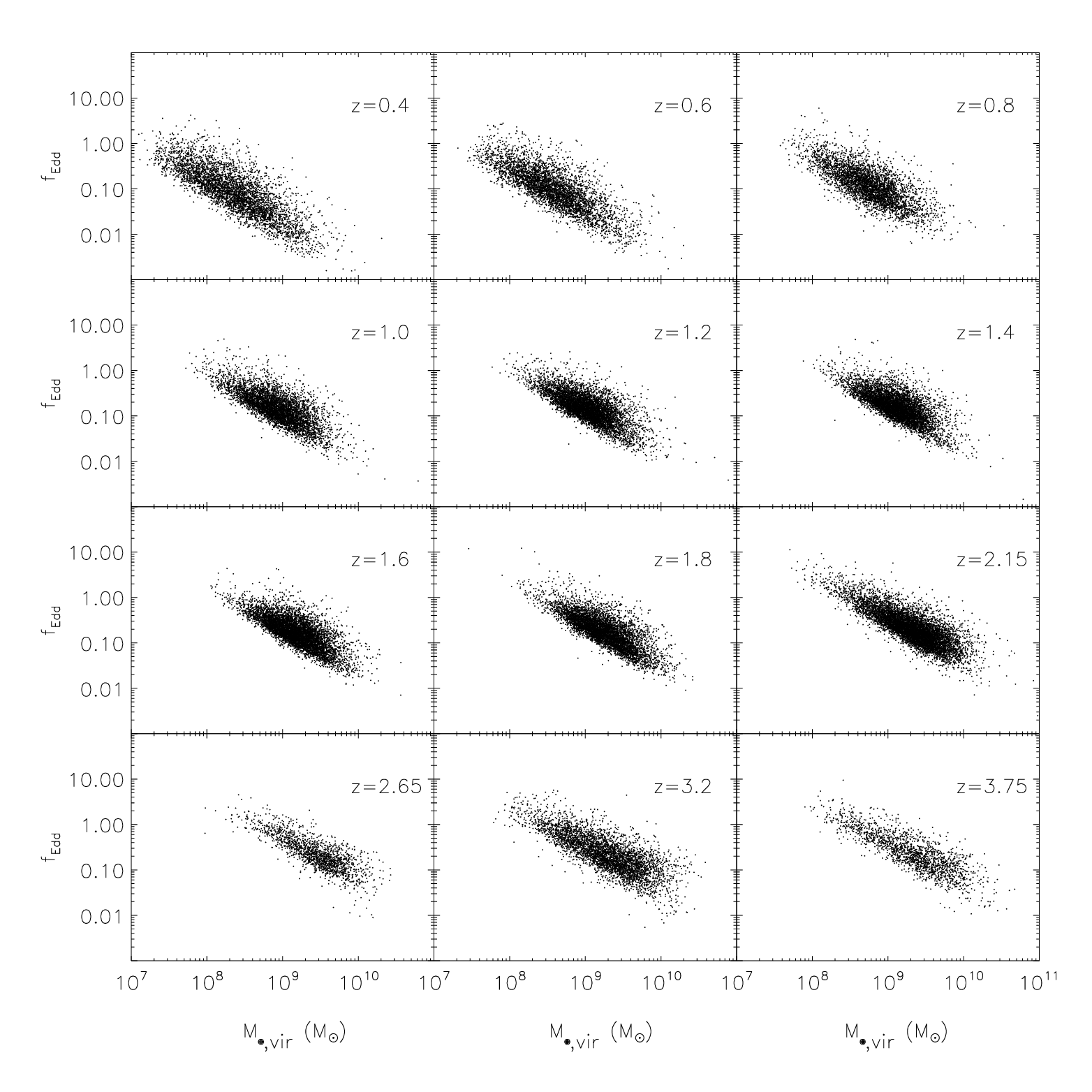}

\caption{Distributions of the SDSS QSOs in the $f_{\rm Edd}- M_{\bullet,
\rm vir}$ plane at different redshift bins. Each grey point represents
an SDSS QSO.  }

\label{fig:f6}
\end{figure*}

The left panel of Figure~\ref{fig:f5} shows the mean value of the
logarithmic radiative efficiency of all the SDSS QSOs in each redshift
bin, and the right panel of Figure~\ref{fig:f5} shows that of the SDSS
QSOs within a given mass range in each redshift bin.  Apparently there
is no significant redshift evolution of the mean logarithmic
efficiency (e.g., $\left<\log\epsilon\right>\sim -0.86$ if adopting
the BB model, and $\sim -0.89$ if adopting the TLUSTY model; see
Table~\ref{tab:t1} and Figure~\ref{fig:f5}). The mean value of $\log
\epsilon$ obtained through the TLUSTY model is slightly smaller than
that obtained through the BB model by $\sim 0.03$~dex. Considering of
the uncertainties in the BC estimations, the mean value of
$\log\epsilon$ shown in the left panel of Figure~\ref{fig:f5} may be
overestimated or underestimated at most by $\sim 0.06$~dex (see
discussions in the end of section~\ref{subsec:lbol}), i.e.,
$\left<\log\epsilon\right> \sim -0.86\pm 0.06$ (or $\sim -0.89\pm
0.06$) by using the BB model (or the TLUSTY model), which corresponds
to a mean efficiency of $\sim 0.14\pm 0.02$ (or $\sim 0.13\pm 0.02$).
Therefore, the mean efficiency of those SDSS QSOs should be in the
range of $\simeq 0.11-0.16$, which is totally independent of but
consistent with those estimations based on the So{\l}tan argument
\citep[e.g.,][]{YT02,YL04, Marconi04, YL08, Shankar09, Zhang12,
Shankar13}. We obtain similar results if alternative using the mean
value of $\epsilon$ or luminosity-weighted $\epsilon$. In the
right panel of Figure~\ref{fig:f5}, we show the mean values of $\log
\epsilon$ for those QSOs in each redshift bin with MBH masses in the
range of $2.5\times 10^8$--$4\times 10^8M_{\odot}$,
$8\times10^8$--$1.3\times 10^9M_{\odot}$ and $2.5\times
10^9$--$4\times 10^9M_{\odot}$, respectively, which are also roughly
consistent with being a constant over the redshift range from $0.3$ to
$4$ (except a seemingly slight increase at $z < 1$ for the QSOs with
MBHs in the range of $2.5\times 10^8$--$4 \times 10^8M_{\odot}$ and
$8\times10^8$--$1.3\times 10^9M_{\odot}$). It appears that the mean
logarithmic efficiency for the QSOs with large MBH masses is larger
than that with small MBH masses (right panel of Figure~\ref{fig:f5}).
However, this tendency seems to be mainly caused by the selection
effects but not intrinsic (as seen also from Figure~\ref{fig:f4} and
discussions in section~\ref{sec:bias}).  

The apparent strong dependence of $\epsilon$ on $\mbhvir $ shown in
Figure~\ref{fig:f4} appears to suggest that more massive MBHs
statistically rotate faster than less massive MBHs in QSOs, as
discussed in \citet{Davis11}. However, it is possible that this
apparent correlation is only a result of the selection effects of the
SDSS QSO sample and the biases induced by various assumptions in the
estimations of $\epsilon$, and the intrinsic radiative efficiency of
individual SDSS QSOs may be not correlated with the MBH true mass
\citep[][but \citealt{Davis11, LD11}]{Raimundo12}. To investigate
whether the above $\epsilon-\mbhvir $ correlation is intrinsic or not,
the key is in understanding the selection effects and the various
biases involved in the $\epsilon$ estimations, which will be studied in
details in the following section~\ref{sec:bias}. 

\subsection{Eddington ratios of SDSS QSOs}

Figure~\ref{fig:f6} shows the distribution of the SDSS QSOs in the
Eddington ratio ($f_{\rm Edd}$) versus the MBH virial mass
($M_{\bullet,\rm vir}$) plane, where $f_{\rm Edd}=L_{\rm bol}/L_{\rm
Edd}(M_{\bullet,\rm vir})$ and $L_{\rm Edd}(M_{\bullet,\rm vir})\sim
1.3\times 10^{46}\ergps (M_{\bullet,\rm vir}/10^8\msun)$. As seen from
Figure~\ref{fig:f6}, the Eddington ratio is strongly anti-correlated
with the MBH virial mass for the SDSS QSOs in every redshift bin,
similar to that found by \citet[][see Figure 13 therein]{Davis11} for
$80$ PG QSOs at low redshift ($z\leq 0.5$). Generally the larger the MBH
virial mass, the smaller the Eddington ratio. However, one should note
that the obtained distribution of the Eddington ratios in the $f_{\rm
Edd}-M_{\bullet,\rm vir}$ plane is also affected by the selection
effects and the biases induced by the usage of the MBH virial mass
(but not the MBH true mass). Similar to that shown for the radiative
efficiency in Figures~\ref{fig:f9} and \ref{fig:f10} in
section~\ref{sec:bias}, the
correlation between Eddington ratio and MBH virial mass can also be
explained by the selection effects and the biases induced by the usage
of the MBH virial masses. In this study, we do not intend to expand
the discussion on Eddington ratios as that done for radiative
efficiencies in section~\ref{sec:bias}.


\section{Biases and Selection Effects in the Radiative Efficiency
Estimations}
\label{sec:bias}

\subsection{Uncertainties in the $\dot{M}\acc$ estimations}
\label{subsec:bias_m}

\begin{figure*}
\centering
\includegraphics[scale=0.55]{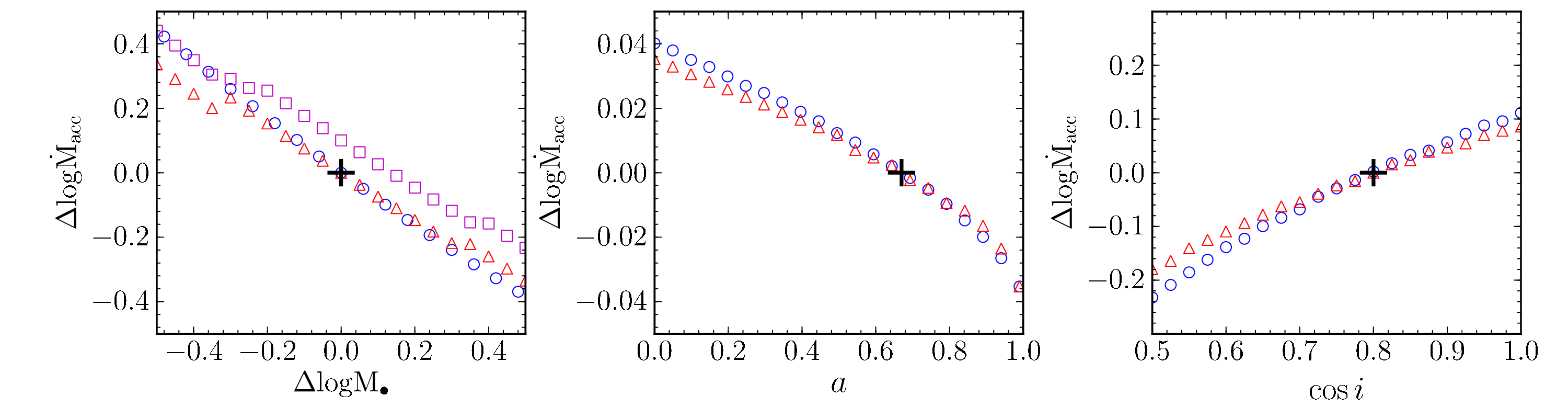}

\caption{Uncertainties in the accretion rate estimations $(\Delta \log
\dot{M}\acc=\log \dot{M}\acc-\log \dot{M}_{\rm acc,t})$ for QSOs with
the same intrinsic parameters $\mbht=10^8 M_{\odot}$, $a=0.67$,
$\dot{M}_{\rm acc,t}=0.67\msunyr$, $\cos i=0.8$ (marked by the black
plus symbol in each panel). Left, middle and right panels show the
uncertainties in the $\dot{M}_{\rm acc}$ estimations solely induced by
the deviation of the virial mass from the true MBH mass ($\Delta \log
M_{\bullet}= \log M_{\bullet,\rm vir}-\log M_{\bullet,\rm t}$), the
set of a fixed value for the MBH spin ($a$), and the set of a fixed
value of the inclination angle ($i$), respectively.  In each panel,
blue circles and red triangles show the uncertainties in the
$\dot{M}\acc$ estimations by adopting the BB model and the TLUSTY
model, respectively. In the left panel,  $\dot{M}_{\rm acc}$ is
obtained by assuming that $a$ and $i$ are the same as the intrinsic
ones while $M_{\bullet,\rm vir}$ is different from $M_{\bullet,\rm
t}$. Magenta squares represent the uncertainties in the $\dot{M}\acc$
estimations obtained from the fitting by adopting the TLUSTY model,
however, the $L_{5100\rm \scriptsize \AA}$ value, which is used for
the model fitting, is generated by the BB model. In the middle panel,
$\dot{M}_{\rm acc}$ is obtained by assuming that $M_{\bullet,\rm vir}$
and $i$ are the same as the intrinsic ones while $a$ is set to be
different from the intrinsic value.  In the right panel, $\dot{M}_{\rm
acc}$ is obtained by assuming that $M_{\bullet,\rm vir}$ and $a$ are
the same as the intrinsic ones while $i$ is set to be different from
the intrinsic value.  }

\label{fig:f7}
\end{figure*}

The above estimations of $\dot{M}\acc$ may be not accurate as (1) the
adopted virial masses of MBHs deviate from the true masses; (2) the
MBH spins and inclination angles are assumed to be fixed values but in
reality they are probably not; and (3) the accretion disk model(s) may
be oversimplified. Figure~\ref{fig:f7} illustrates the uncertainties
in the $\dot{M}_{\rm acc}$ estimations induced by these various
factors as detailed below.

First, the uncertainty due to the usage of the virial mass as the true
MBH mass. The estimated virial masses of MBHs may deviate from the
true masses with a scatter of $0.3-0.4$~dex
\citep[e.g.,][]{Shen08,Shen11}. We perform following calculations to
quantify the uncertainty of the $\dot{M}\acc$ estimation induced by
the usage of the virial mass (rather than the true MBH mass).
Assuming a typical QSO, of which the true MBH mass is
$\mbht=10^8\msun$ and spin is $a=0.67$, accreting material via a true
rate of $\dot{M}_{\rm acc,t}=0.67 \msunyr$ (corresponding to an
Eddington ratio of $0.3$), and the inclination angle of the disk to
the line of sight is $i=$acos$(0.8)$.  The optical band luminosity at
$5100$\AA~ of this system is predicted to be $L_{5100\rm \scriptsize
\AA}=10^{44.56}\ergps$ by adopting the BB model.  Assuming a virial
mass of the MBH $\mbhvir $, scattered around $\mbht$, we may estimate
the accretion rate $\dot{M}\acc$ according to $L_{5100\rm \scriptsize
\AA}$ and $\mbhvir $ through the procedures described in
section~\ref{subsec:mdot}. For this specific case, $5100$\AA,
$3000$\AA and $1350$\AA~ are all at 
the left side of the turnover of the disk emission spectrum (which 
is also true for more than $90\%$ of the type 1 SDSS QSOs).
Clearly, $\dot{M}\acc$ can be
underestimated (or overestimated) by adopting the BB model if
$\mbhvir$ is larger (or smaller) than $\mbht$ (see the blue circles in
the left panel of Figure~\ref{fig:f7}). For a typical deviation of
$\mbhvir $ from $\mbht$, i.e., $0.3$~dex, $\dot{M}\acc$ may be either
underestimated or overestimated by a factor of $\sim 2$.  If
alternatively adopting the TLUSTY model for the same system, we have
$L_{5100\rm \scriptsize \AA}=10^{44.51}\ergps$. Similarly we can
obtain the best fit of $\dot{M}\acc $ for each set of ($L_{5100\rm
\scriptsize \AA}$, $\mbhvir $), as represented by the red triangles in
the left panel of Figure~\ref{fig:f7}.  In these cases, the
uncertainties in the estimated $\dot{M}\acc$ are similar to that by
the BB model above. For an observational sample of QSOs, the virial
masses of the QSO central MBHs are more likely to be overestimated (or
underestimated) at the high (or low) mass end, which may consequently
leads to the overestimate (or underestimate) of the radiative
efficiencies and thus could contribute to the apparent correlation
found in section~\ref{sec:estepsilon} (see also the discussions in
\citealt{Davis11}). However, we note here that the turnover of the
disk emission spectra may move to the left side of $1350$\AA~ for
some QSOs with extreme large MBH masses ($\ga 10^{10}\msun$) and small
Eddington ratios ($\la 0.04$). For these rare cases, if the virial masses
of the QSO central MBHs are overestimated, the radiative efficiencies
may be consequently underestimated, which is totally different from 
that for QSOs with smaller MBH masses and high Eddington ratios.
 
Second, the uncertainties due to the usage of a fixed value of $a$ and
a fixed value of $i$.  We first assume that the true MBH mass
$M_{\bullet,\rm t}$ and inclination angle $i$ of the systems are known
and arbitrarily adopt a value of $a$ to do the fitting in order to
check the uncertainty solely due to the set of a fixed $a$ (different
from the intrinsic value).  We find that the best-fit of $\dot{M}\acc$
only deviates slightly, at most $0.04$~dex or $10\%$, from its true
value (see the middle panel of Figure~\ref{fig:f7}).  The reason is
that $L_{\rm \scriptsize 5100\AA}$ is insensitive to the MBH spin at
$\mbht \la 10^{8}M_{\odot}$ in the thin accretion disk model(s) for
QSOs.  If choosing larger $\mbht $ (e.g., $\ga 10^9M_{\odot}$), the
uncertainty introduced to the estimate of $\dot{M}\acc$ is no more
than $50$\% by an arbitrary set of $a$ \citep[see also][]{Davis11}.
Most SDSS QSOs are type 1 QSOs, their inclination angles $i$ may be
uniformly distributed in $\cos i$ between $0.5$ and $1$.  We now
assume that the true MBH mass $M_{\bullet,\rm t}$ and spin $a$ of the
systems are known and obtain the best fit values of $\dot{M}\acc$ by
arbitrarily setting the value of $i$. The uncertainty induced by the
set of a fixed value for $i$ ($\cos i$) is shown in the right panel of
Figure~\ref{fig:f7}, which are at most $0.2$~dex.  In this section, we
set a fixed value of $i=$acos$(0.8)$ for the fitting, with which the
results obtained here can be directly compared with that found in
\citet{Davis11} and \citet{Raimundo12}.

Third, the uncertainty induced by the choice of a specific accretion
disk model. The real accretion process may be not accurately reflected
by either the BB model or the TLUSTY model. Adopting a specific disk
model may then introduce some (systematic) bias to the accretion rate
estimations. To illustrate this uncertainty, we first generate a number
of QSOs with optical luminosity predicted by the TLUSTY (or BB) model,
but then adopt the BB (or TLUSTY) model to estimate $\dot{M}\acc $ by
using $L_{5100\rm \scriptsize \AA}$ and $\mbhvir$. We find that the
estimated $\dot{M}\acc$ could be systematically under-estimated or
(over-estimated) due to the choice of an accretion disk model that not
exactly reflect the underlying true disk accretion physics (for
example, see the magenta squares shown in the left panel of
Figure~\ref{fig:f7}). 

\begin{figure*}
\centering
\includegraphics[scale=0.55]{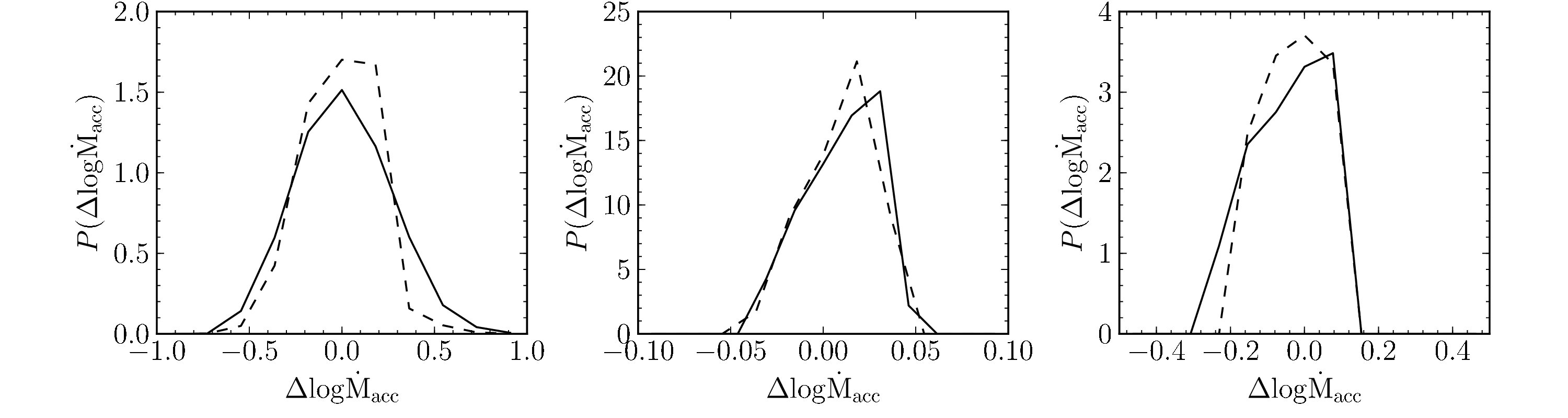}

\caption{Probability distributions of $\Delta \log \dot{M}\acc =\log
\dot{M}\acc -\log \dot{M}_{\rm acc,t}$ obtained from the mock QSO
samples by adopting different assumptions in the fitting.  The first
mock sample is generated by assuming all QSOs have the same intrinsic
properties ($\mbht =10^8M_{\odot}$, $a=0.67$ and $\dot{M}_{\rm acc,t}=0.67
\msunyr$ and $i=$acos$(0.8)$). The QSOs in the second and third mock
samples have the same intrinsic properties as that in the first mock
sample, except that $a$ is assumed to be distributed uniformly in the
range from $0$ to $0.99$ and $i$ is assumed to be uniformly
distributed in $\cos i$ between $0.5$ and $1$, respectively. The
virial mass is randomly assigned to each mock QSO in the first sample
according to a Gaussian distribution with a mean of $\mbht $ and a
scatter of $0.3$~dex. Left panel shows the resulted distribution from
the estimations of $\dot{M}\acc$ for the first mock sample by assuming
that the virial mass is the ``true'' MBH mass, $a=0.67$,
and $i=$acos$(0.8)$. Middle and right
panels show the resulted distributions for the second sample and the
third sample, respectively, by assuming $M_{\bullet}=M_{\bullet,\rm t}$, $a=0.67$ 
and $i=$acos$(0.8)$. In each panel, the solid and dashed
lines represent the results obtained by adopting the BB model and the
TLUSTY model, respectively.  }

\label{fig:f8}
\end{figure*}

We further perform Monte-Carlo calculations to illustrate the
uncertainties in the accretion rate estimations due to the above
factors. We first generate mock QSO samples in which all QSOs have the
same intrinsic properties ($\mbht =10^8\msun $ and $\dot{M}\acc =
0.67\msunyr$). The spin and inclination angle are set to be $a=0.67$
and $i=$acos$(0.8)$, respectively, for all the QSOs in the first
sample; while $i=$ acos$(0.8)$ and $a$ is set to be uniformly
distributed in the range from $0$ to $0.99$ for the QSOs in the second
sample; and $a=0.67$ and  $i$ is assumed to be uniformly distributed
in $\cos i$ between $0.5$ and $1$ for the QSOs in the third sample. We
randomly assign $\mbhvir $ to each mock QSO in the first sample
according to a Gaussian distribution around $\mbht$ with a scatter of
$0.3$~dex. We obtain $L_{5100\rm \scriptsize \AA}$ by both the BB
model and the TLUSTY model for each QSO, respectively.  We obtain
three samples of QSOs with given observational properties of
$L_{5100\rm \scriptsize \AA}$ and $\mbhvir $ (or $M_{\bullet,\rm t}$).
We adopt the procedures described in section~\ref{subsec:mdot} to
estimate $\dot{M}\acc$ for these mock QSOs. Figure~\ref{fig:f8} shows
the probability distribution of $\Delta \log \dot{M}\acc =\log
\dot{M}\acc -\log \dot{M}_{\rm acc,t}$ for the first (left panel), 
second (middle panel) and third sample (right panel),
respectively. In each panel, the solid and dashed lines show the
results obtained by adopting the BB model and the TLUSTY model,
respectively. In the left panel, $\dot{M}\acc$ are obtained for each
QSO in the first sample by assuming $a=0.67$, $\cos i =0.8$, and the
virial mass as the ``true'' MBH mass; in the middle and right panel,
$\dot{M}\acc$ is obtained for each QSO in the second and third samples
by assuming $M_{\bullet,\rm vir}=M_{\bullet,\rm t}$, $\cos i=0.8$ and
$a=0.67$. Figure~\ref{fig:f8} clearly illustrates that
the uncertainties of the $\dot{M}\acc $ estimations induced by the
deviations of the virial masses of MBHs from the true masses could be
as large as a factor of $2$ and can have significant effects on the
$\epsilon$ estimations, while the errors due to inaccurate settings of
$a$, $i$ and the choice of the disk model are relatively less
significant.  If alternatively choosing other values of $\mbht$ and
$\dot{M}\acc$, we obtain similar results as that shown in
Figure~\ref{fig:f8}.

\subsection{Monte-Carlo simulations }
\label{sec:MCsim}

In this section, we investigate whether the correlation between
radiative efficiency and MBH mass is intrinsic or simply due to
various biases in the accretion rate estimations and/or the selection
effects of the SDSS QSO sample (mainly a flux limit sample). 

First, we generate mock SDSS QSOs based on the true mass function of
MBHs (BHMF, $\Phi(\bh)$) and the true Eddington ratio ($f_{\rm Edd}$)
distribution function (ERDF, $\Phi(f_{\rm Edd}$)) obtained by
\citet[][see their Tables 1 and 2]{Kelly13}. According to the BHMF and
the ERDF at each redshift bin, we randomly assign the MBH mass $\mbht
$ (in the range from $10^6\msun - 10^{11}\msun$) and the Eddington
ratio $f_{\rm Edd}$ (in the range from $10^{-4}$ to $10^{1}$) to each
mock QSO.  \citet{Kelly13} argued that the Eddington ratio
distribution may be independent of the MBH mass and the difference in
the Eddington ratio distribution for MBHs with different masses at
$0.8<z<2.65$ might be only an effect of the uncertainties in the MBH
mass estimations by the usage of the CIV lines. We assume that the 
Eddington distribution is the same for MBHs with different masses.
Assuming that $\epsilon_{\rm t}$ is randomly distributed over the
range from $0.057$ to $0.31$, corresponding to a spin range from $0$
to $0.998$,\footnote{If alternatively assuming that the radiative
efficiencies for all QSOs are the same, i.e., $\epsilon_{\rm t}=0.13$
(similar to the mean efficiency shown in Figure~\ref{fig:f3};
or $0.1$ or $0.2$), corresponding to a spin $a\sim 0.83$ (or $\sim 0.67$
or $0.96$),  we
find little changes of our following conclusions.} the intrinsic
accretion rate is then $\dot{M}_{\rm acc,t}= f_{\rm Edd} L_{\rm
Edd}/\epsilon_{\rm t}c^2$. The inclination angle $i$ is assumed to be
uniformly distributed in $\cos i$ from $0.5$ to $1$. Given the
intrinsic parameters of each mock QSO, i.e., $\mbht $, $a$,
$\dot{M}_{\rm acc,t}$ and $i$, the optical band luminosity $L_{5100\rm
\scriptsize \AA}$ can be obtained by adopting the BB (or TLUSTY)
model. A virial mass of the central MBH in each mock QSO is assigned
according to a Gaussian distribution with a mean of $\log \mbht $ and
a scatter of $0.3$~dex. With the above procedure, we generate $5\times
10^6$ mock QSOs in each redshift bin, and finally we have a large
number of mock QSOs that have not only the ``observational
measurements'' of their virial mass $M_{\bullet,\rm vir}$ and the
optical band luminosity(/-ies) $L_{5100\rm \scriptsize \AA}$ (and/or
$L_{3000\rm \scriptsize \AA}$, and/or $L_{1350\rm \scriptsize \AA}$)
but also
known intrinsic properties of $\mbht$, $\dot{M}\acc $ and $\epsilon$.
The optical band luminosities are produced at $5100$\AA~ for those
QSOs with $z<0.35$, at $5100$\AA~ and $3000$\AA~ for those QSOs with
$0.35<z<0.9$, at $3000$\AA~ for those QSOs with $0.9<z<1.5$, at
$3000$\AA~ and $1350$\AA~ with $1.5<z<2.25$, and at $1350$\AA~ for
those QSOs with $2.25<z<4$, respectively, in order to mimic the SDSS
QSO sample.  We take these mock QSOs as the parent populations of the
QSOs in each redshift bin that can be detected by a survey like the
SDSS.  

With the ``observational'' properties of the mock QSOs, i.e., $\mbhvir
$ and $L_{5100\rm \scriptsize \AA}$ (and/or $L_{3000\rm \scriptsize
\AA}$, and/or $L_{1350\rm \scriptsize \AA}$), we can adopt the same
procedure as that described in section~\ref{sec:estepsilon} to
estimate $\epsilon$. In order to single out the effects on the
$\epsilon$ estimations due to the deviation of $\mbhvir $ from $\mbht$,
we also estimate $\epsilon$ for the mock QSOs by using their optical
luminosity $L_{5100\rm \scriptsize \AA}$ (and/or $L_{3000\rm
\scriptsize \AA}$, and/or $L_{1350\rm \scriptsize \AA}$) and the true
MBH mass $\mbht $. We further consider the selection criteria for
those mock QSOs similar to that for the SDSS QSOs, i.e., only those
mock QSOs with apparent $i$-band magnitude $\leq 19.1$ at $z\leq 2.9$
and $\leq 20.2$ at $z>2.9$ can be taken as the mock SDSS QSOs (see
section~\ref{sec:SDSSQSO}). To select mock SDSS QSOs, we first
convert $L_{5100\rm \scriptsize \AA}$ (or $L_{3000\rm \scriptsize
\AA}$, or $L_{1350\rm \scriptsize \AA}$) to the monochromatic
luminosity at $i$-band ($7471$\AA) by assuming a power law spectrum
with a canonical slope of $\alpha_{\rm opt}=-0.5$, and then use the
$K$-correction, including the effects of both the continuum and the
emission lines, given by \citet{Richards06b} to calculate the apparent
$i$-band magnitude of each mock QSO. Here we adopt $\alpha_{\rm
opt}=-0.5$ because the K-corrections given by \citet{Richards06b} are
obtained by using a canonical slope of $-0.5$.  Finally, we obtain the
mock SDSS QSOs according to the above selection criteria, which are
consist of a small fraction ($\sim 1-5\%$) of the parent mock sample at each
redshift bin. Here we
adopt the standard $\Lambda$CDM cosmology with $(H_0, \Omega_{\rm M},
\Omega_{\Lambda}) =71\kms {\rm Mpc}^{-1}, 0.27, 0.73$).  Similar to
that in section~\ref{sec:estepsilon}, we do the simulations by adopting both the BB model
and the TLUSTY model, and we find that the results obtained from two
different models are similar. For simplicity, only the results
obtained by adopting the BB model are presented below in this section. 

\begin{figure*}
\centering
\includegraphics[scale=0.95]{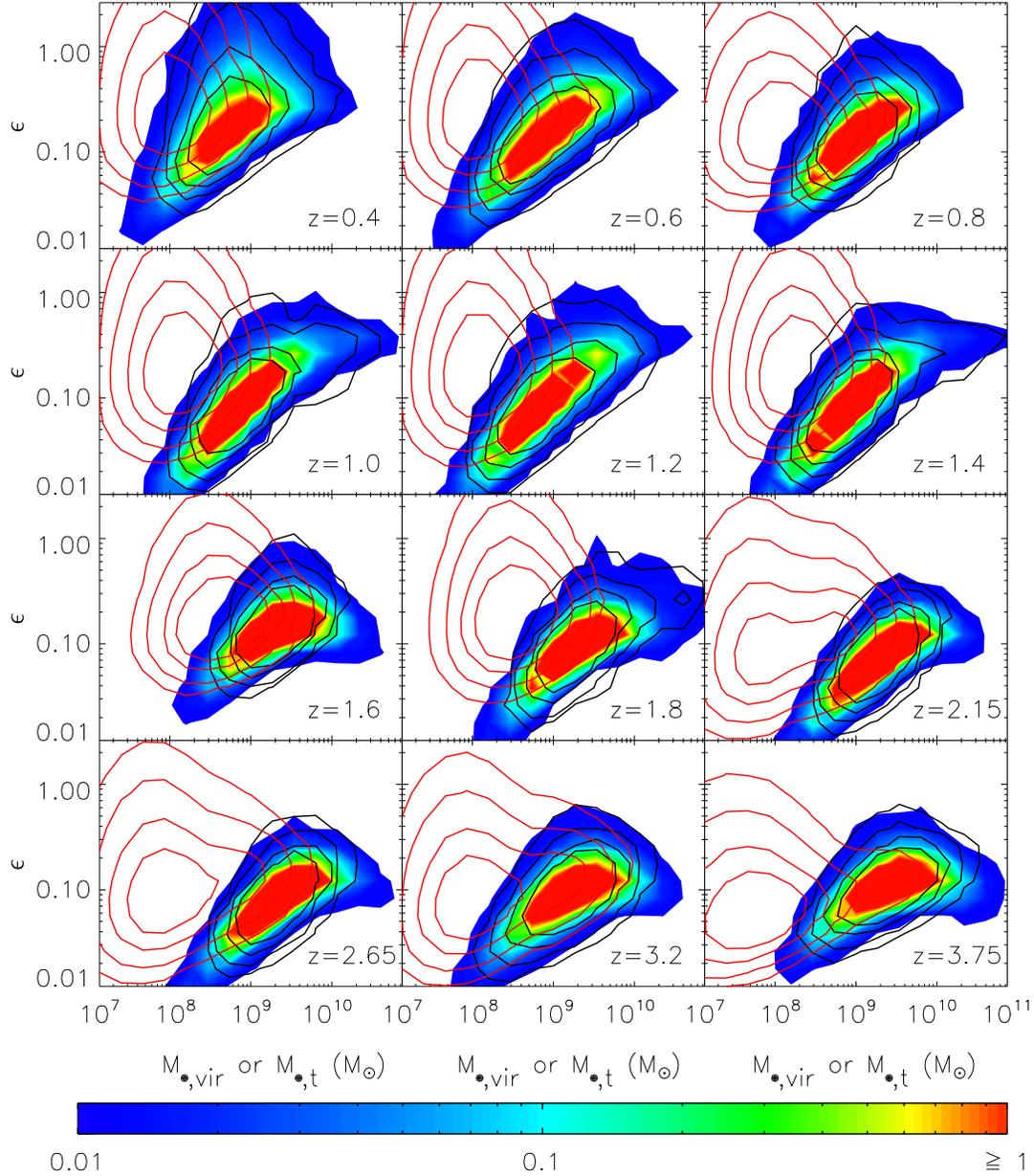}

\caption{Probability density distributions of the mock QSOs at
different redshift bins in the $\log \epsilon$--$\mbhvir $ (or $\log
\epsilon$--$\mbht $) plane. In each redshift bin, the red contour
lines represent the probability density distribution of all the mock
QSOs without considering the selection effect, and the radiative
efficiency $\epsilon$ of each mock QSO is estimated by using
($L_{5100\rm \scriptsize \AA}$, $\mbht $); the black contour lines
represent the distribution of those mock QSOs that satisfy the
selection criteria as that for the SDSS QSOs, and the radiative
efficiency $\epsilon$ of each mock QSO is estimated by using
($L_{5100\rm \scriptsize \AA}$, $\mbht $); and the filled color
contours represent the distribution of those mock QSOs that satisfy
the selection criteria as that for the SDSS QSOs, and the radiative
efficiency $\epsilon$ of each mock QSO is estimated by using
($L_{5100\rm \scriptsize \AA}$, $\mbhvir $). The probability densities
represented by the four contour lines (both in black and in red) from
the the outer to inner regions are $0.01$, $0.05$, $0.2$ and $0.5$,
respectively, in units of per $\log M_{\bullet,\rm vir}$ (or $\log
M_{\bullet,\rm t}$) per $\log \epsilon$. The bar at the bottom shows
the scales of the filled color contours.  }

\label{fig:f9}
\end{figure*}

\begin{figure*}
\centering
\includegraphics[scale=0.95]{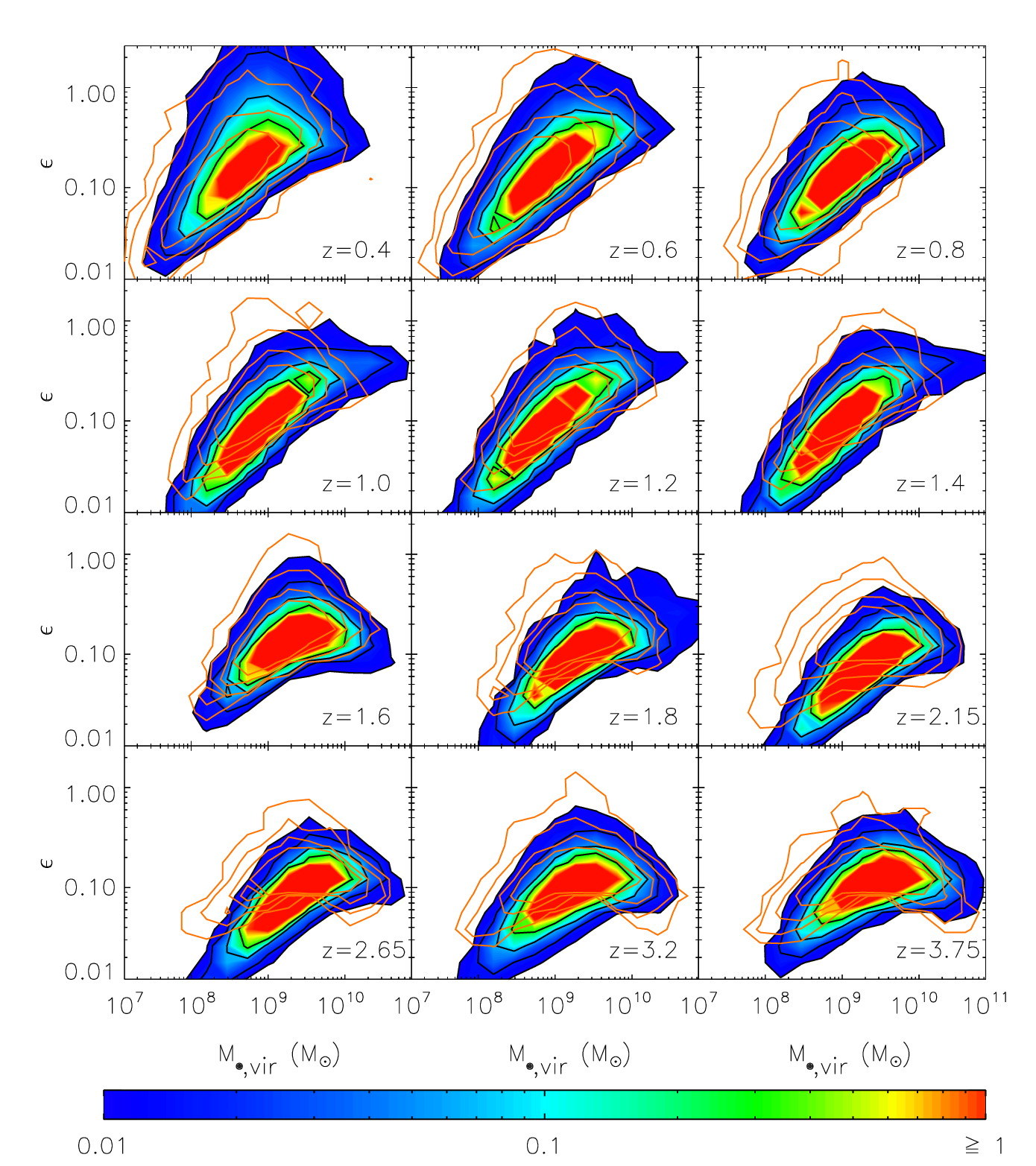}

\caption{Probability density distributions of the mock SDSS QSOs and
the SDSS QSOs in the $\log\epsilon$--$\log \mbhvir $ plane. The filled
color contours (and also the black contour lines) represent the
distribution of the mock SDSS QSOs (the same as the filled color
contours in Figure~\ref{fig:f9}) and the orange contour lines
represent the distribution of the SDSS QSOs in each redshift bin. The
SDSS QSOs here correspond to the observational SDSS QSO sample (see
section~\ref{sec:SDSSQSO} and Figure~\ref{fig:f4}). The probability
densities represented by the four contour lines (both in black and in
orange) from the the outer to inner regions are $0.01$, $0.05$, $0.2$
and $0.5$, respectively, in units of per $\log M_{\bullet,\rm vir}$
per $\log \epsilon$. The bar at the bottom shows the scales of the
filled color contours.  }

\label{fig:f10}
\end{figure*}

Figure~\ref{fig:f9} shows the normalised probability density 
distribution of the mock QSOs on the $\log\epsilon$--$\mbhvir$ (or
$\log\epsilon$--$\mbht$) plane. Without considering the selection
effects, the radiative efficiencies $\epsilon$ of the mock QSOs
estimated from $L_{5100\scriptsize \rm \AA}$ and $\mbht$ appear to be
independent of $\mbht $ (see the red contour lines in each redshift
bin). By performing Spearman's rank correlation analysis, we do not
find any strong correlations between $\epsilon$ and $\mbht $ for the
whole mock sample (without considering the selection effects). The
estimated $\epsilon$ scatter around the mean of the input values $\sim
0.18$ mainly because the BC is randomly assigned to each mock QSO
according to the empirical SED of QSOs as described in
section~\ref{subsec:lbol}.\footnote{The contours center seem to be not
exactly the mean value ($0.18$) of the input efficiencies for some
redshift bins mainly because of the following reason. The
efficiency for each QSO is estimated by using the bolometric
luminosity obtained empirically and the accretion rates obtained by
adopting the standard thin disk model. This method is not completely
self-consistent, i.e., the efficiency estimate could be slightly
offset from the input intrinsic efficiency for the standard thin disk
model if the empirical SED is not exactly the same as the SED
predicted by the standard thin disk model, which is a caveat of
both our model and other previous models \citep[e.g.,][]{Davis11}.
} For the mock SDSS QSOs, there is an obvious strong correlation
between the radiative efficiencies $\epsilon$ (estimated from
$L_{5100\rm \scriptsize \AA}$ and $\mbht $) and the true MBH mass
$\mbht $ (see the black contour lines in each panel of
Figure~\ref{fig:f9}). Similar to that for the SDSS QSOs, we adopt a
power law form to fit these correlations, i.e., $\epsilon\propto
M^{\beta}_{\bullet,\rm t}$, and find $\beta =0.46, 0.50, 0.51, 0.56,
0.57, 0.56, 0.48, 0.57, 0.63, 0.60, 0.45$ and $0.36$ for the redshift
bins from low to high, respectively. Further considering the
deviation of $\mbhvir $ from $\mbht $, we find that the radiative
efficiency $\epsilon$ (estimated from $L_{5100\rm \scriptsize \AA}$
and $\mbhvir $) are strongly correlated with $\mbhvir $ (see the
filled color contours in Figure~\ref{fig:f9}).  Adopting the same
power-law form to fit these correlations,  i.e., $\epsilon \propto
M^{\beta}_{\bullet, \rm vir}$, we find $\beta= 0.54, 0.60, 0.60, 0.67,
0.68, 0.69, 0.33, 0.48, 0.56, 0.52, 0.37$ and $0.28$ for the redshift
bins from low to high, respectively. It appears that the slope $\beta$
is larger at low redshift ($z\la 1.8$) and smaller at high redshift
($z\ga 2$), compared with the cases without considering the
deviation of $\mbhvir $ from $\mbht $, which is mainly due to the
biases induced by the deviations of the virial masses from the true
MBH masses. Similar to that for the SDSS QSOs, the slope $\beta$ is
relatively large at low redshift and small at high redshift.

Figure~\ref{fig:f10} shows the comparison of the normalised probability density
distributions of the mock SDSS QSOs with that of the SDSS QSOs in the
$\log\epsilon$--$\mbhvir $ plane. Apparently, the distributions of the
mock SDSS QSOs in the $\log\epsilon$--$\log \mbhvir $ plane are
roughly consistent with that of the SDSS QSOs in all the redshift
bins. According to Figures~\ref{fig:f9} and \ref{fig:f10}, it is clear
that the selection effects can be the dominant factors that cause the
strong correlation between $\epsilon$ and $\mbhvir $, and the usage of
the virial mass $\mbhvir $ as the true MBH mass also contribute some
to the correlation but it is less significant comparing with the
selection effects. As discussed in section~\ref{subsec:bias_m}, the
uncertainties in the $\epsilon$ estimations induced by the usage of a
constant inclination angle $i$ play little role in generating the
$\epsilon$--$ \mbhvir $ correlation.

The simplest explanation for the arouse of an apparent correlation
between $\epsilon$ and $M_{\bullet,\rm vir}$, as an addition to the
above Monte-Carlo analysis, is as follows. In the standard thin
accretion disk scenario, the accretion rate is related to the optical
luminosity as $\dot{M}_{\rm acc} \propto L^{3/2}_{\rm opt}
M^{-1}_{\bullet}$ according to \citet{Davis11}. Since $L_{\rm bol}
\propto L_{\rm opt}$, $\epsilon = L_{\rm bol}/\dot{M}_{\rm acc} c^2
\propto L^{-1/2}_{\rm opt} M_{\bullet}$. For a given $M_{\bullet}$,
the larger the efficiency, the smaller the optical luminosity $L_{\rm
opt}$ and thus the larger the probability that the QSO is missed in a
flux limited sample. The smaller the MBH mass, the larger the
probability that QSOs with high radiative efficiencies can be detected
by a flux limited survey like SDSS. For high redshift bins ($z\ga
2.15$), the general tendency of increasing efficiency with increasing
MBH mass is similar to that for low redshift bins if the MBH mass is
smaller than a few times $10^9M_{\odot}$; while it turns over if the
MBH mass is larger. For the QSOs with MBH masses larger than a few
time $10^9\msun$ at high redshift $z\geq 2.25$, the observational
measurement is only provided at $1350$\AA, which is probably at the
right side of the disk emission spectrum. For these cases, the
empirical SEDs adopted in this study, a monotonically increasing
function at $1\mu$m$-1300$\AA, are somewhat different from the disk
emission spectra, which might lead to underestimation of the
bolometric luminosities and thus underestimation of the radiative
efficiencies. This could be part of the reason that the estimated
radiative efficiencies of many QSOs with MBH mass $\ga
10^{10}M_{\odot}$ at high redshift bins ($z\ga 2.15$) deviate from the
correlation between efficiency and MBH mass at lower redshift.
Furthermore, the accretion rate could be overestimated by using the
monochromatic luminosity $L_{1350\rm \scriptsize \AA}$ if the MBH
virial mass is an overestimate of the true MBH mass (probably true for
many QSOs at the high mass end at high redshift, e.g., $z\ga 2.25$)
since the turnover of the disk emission spectra may move to the left
side of $1350$\AA~ for some QSOs with extreme large MBH mass ($\ga
10^{10} M_{\odot}$) and small Eddington ratio ($\la 0.04$; see
discussion in section~\ref{subsec:bias_m}). Such a behavior is quite different from
that for the cases with lower MBH masses and higher Eddington ratios.
The turnover of the trend for the QSO radiative efficiencies in high
redshift bins ($z\ga 2.15$) at high mass end can also be partly
explained by this.

One may note, however, that the probability distributions of the mock
SDSS QSOs in the $\log\epsilon$--$\log \mbhvir $ plane seem to be
offset from that of the SDSS QSOs. The slopes $\beta$ of the best fits
are also not exactly the same as that of the SDSS QSOs.  Especially in
the high redshift bins ($z\ga 2$), the $\epsilon$--$\mbhvir$
correlation obtained from the mock QSOs is still significant, while
that obtained from the SDSS QSOs is very weak. The above differences
between the mock SDSS QSOs and the SDSS QSOs might be not a surprise
as a number of complications are not considered in the above
$\epsilon$ estimations for both the SDSS QSOs and the mock QSOs.
First, we do not consider the contamination by the host galaxy light
and the dust extinction for each individual SDSS QSO, which may
introduce errors to both the $\dot{M}\acc $ and $L\bol$ estimations
and consequently the $\epsilon$ estimations for the SDSS QSOs (see
also \citealt{Raimundo12}).  Second, the bolometric luminosity $L_{\rm
bol}$ is obtained according to the empirical SED of QSOs and its
associated scatters, which are obtained from the multi-wavelength
observations of samples with a small number of QSOs that only cover
limited ranges of luminosity and redshift (see
section~\ref{subsec:lbol}). The usage of this SED may introduce not
only scatters but also systematic biases in the $\epsilon$
estimations.  The small spread in the shapes of the adopted empirical
SEDs may be also important for the resulted statistics of the
$\epsilon$ estimations, as pointed out by \citet{LD11}. Third, the
mock QSOs are generated according to the true BHMF and ERDF given by
\citet{Kelly13}. In \citet{Kelly13}, the obtained BHMF and ERDF can
best fit the QSO observations, however, they are obtained by assuming
a constant bolometric correction. Since the BCs of the type 1 QSOs may
have a large scatter and may also depend on the Eddington ratio and
the true MBH mass \citep[e.g.,][]{Vasudevan07, Vasudevan09, Lusso10,
JinDone12}, the BHMF and ERDF obtained by \citet{Kelly13} may still
deviate somewhat from the true distributions, which may further
introduce some uncertainties in the distributions of the mock SDSS
QSOs in the $\log\epsilon$--$\log \mbhvir $ plane.  One needs to
self-consistently consider all the factors, including the BHMF, ERDF,
BCs and the radiative efficiencies, etc., simultaneously, to fully
solve this problem..

In principle, one could also check whether can an intrinsic
correlation between $\epsilon_{\rm t}$ and $\mbht$ be ruled out (or
confirmed) by comparing the results obtained from the SDSS QSOs with
that obtained from mock SDSS QSO samples. To check this, we
re-generate the mock SDSS QSO sample by assuming that $\epsilon_{\rm
t}=0.1 (\mbht /10^8\msun)^{0.5}$, as an alternative to the initial
setting of a random $\epsilon_{\rm t}$ in the above calculations.
According to this new mock SDSS QSO sample, we find that the resulted
correlation between $\epsilon$ and $\mbhvir$ is generally steeper than
that obtained above by assuming a constant $\epsilon_{\rm t}$ in each
redshift bin. For example, $\beta=0.56, 0.62$, and $0.75$ for the
first three redshift bins, respectively. The differences in $\beta$
may help to distinguish whether there is indeed an intrinsic
correlation between $\epsilon_{\rm t}$ and $\mbht$. Because of the
various complications and large uncertainties in the $\epsilon$
estimations (see discussions in the above paragraph), however, it is
still difficult to fully rule out the existence of an intrinsic
$\epsilon_{\rm t}$--$\mbht$ correlation just by comparing the results
obtained from the SDSS QSOs with that obtained from such mock samples.
We conclude that the intrinsic correlation between the radiative
efficiency and the true MBH mass, if any, must be much weaker than the
apparent correlation between $\epsilon$ and $\mbhvir$ found for the
SDSS sample in each redshift bin.

To close this section, we conclude that the strong correlations
between $\epsilon$ and $\mbhvir $ found in
section~\ref{sec:estepsilon} can be produced by and mainly due to the
selection effects of the SDSS QSO sample and the biases in the
$\epsilon$ estimations induced by the usage of $\mbhvir $ as the true
MBH mass. By studying a number of QSOs (or AGNs) with different
luminosities and MBH masses and redshifts, \citet{Raimundo12} find
that the $\epsilon$--$\mbhvir $ relation found by \citet{Davis11} may
be an artifact of the small parameter space covered by the DL sample.
With the large data set of the SDSS QSO catalog, we have demonstrated
that a correlation between $\epsilon$--$M_{\bullet,\rm vir}$ can be
generated by mocking SDSS QSOs through modelling various selection
effects and biases (though there is no input intrinsic correlation
between $\epsilon$--$\mbht $), and our results are consistent with
that of \citet{Raimundo12}. Note that the empirical SED of QSOs, more
or less uniform, is adopted to calculate the bolometric luminosity of
each SDSS QSO in section~\ref{subsec:lbol}.  As argued in
\citet{LD11}, the small spread in the SED shape is an intriguing
question that currently does not have a simple answer.

\section{Conclusions}
\label{sec:conclusions}

Radiative efficiencies of the disk accretion processes in individual
QSOs are related to the spins of the central MBHs, which may be
profoundly connected to the MBH assembly history as suggested by a
number of recent studies \citep[e.g.,][]{Gammie04,Volonteri05,King06,
King08,Berti08, Dubois13}. It is therefore of great importance to
estimate the radiative efficiency of individual QSOs and investigate
its statistical distribution among QSOs. In this study, we estimate
the radiative efficiency individually for a large number of SDSS QSOs.
We first estimate the accretion rate for each QSO by matching the
detected optical band luminosity(/-ies) with that predicted by the
disk model, by adopting the thin disk accretion model and assuming
that the true mass of the central MBH is the same as that given by the
virial mass estimator(s).  We also estimate the bolometric luminosity
of each QSO by adopting the empirical spectral energy distribution
suggested by various multi-wavelength observations of small QSO
samples. With the estimated accretion rate and the bolometric
luminosity, we obtain the radiative efficiency for each SDSS QSO.  We
find an apparent strong correlation between the radiative efficiency
and the MBH virial mass in low redshift bins and it becomes weak in
high redshift bins.  In the lowest redshift bin ($0.3\leq z<0.5$),
this apparent correlation ($\epsilon\propto M^{\beta}_{\bullet, \rm
vir}$, and $\beta\sim 0.56-0.64$) is roughly consistent with that
found by \citet{Davis11} for the DL QSOs in a similar redshift range.
We also find that the mean radiative efficiencies of the SDSS QSOs
are consistent with being a constant $\simeq 0.11-0.16$ (though with
large scatters) over the redshift range from $0.3$ to $4$, which does
not suggest any significant evolution with redshift. This estimate of
the mean radiative efficiency of QSOs is totally independent of but
roughly consistent with those estimations based on the So{\l}tan
argument \citep[e.g.,][]{YT02,YL04,Marconi04, YL08, Shankar09,
Zhang12, Shankar13}.

With the enormous large sample of the SDSS QSOs, it is possible to
statistically model the various biases in the estimations of the
radiative efficiency and the selection effects of the SDSS QSOs. To do
so, we generate mock SDSS QSO samples according to the true MBH mass
function and the Eddington ratio distribution obtained in
\citet{Kelly13}, by involving the selection criteria of the SDSS QSOs
and the uncertainties in the MBH virial mass estimations and the
inclination angles. We estimate the radiative efficiency for QSOs in
each mock sample  by adopting the same method as that for the SDSS QSO
and obtain the probability density distribution of those mock QSOs in
the radiative efficiency versus the MBH virial mass plane. We find
that the $\epsilon$--$\mbhvir $ correlations for the SDSS QSOs in
different redshift bins can be well explained by the selection effects
and the biases induced by the usage of $\mbhvir $ as the true MBH
mass, and the selection effects play the dominant roles in leading to
the $\epsilon$--$\mbhvir $ correlation, as suggested by
\citet{Raimundo12}. We conclude that the current SDSS QSO data is
consistent with no intrinsic correlation between the QSO radiative
efficiency and the true MBH mass. 

In principle, the accretion rate of a QSO may be better determined by
fitting the QSO spectrum covering a wide range of wavelengths (e.g.,
from the infrared band to the hard X-ray band), through an elaborate
accretion disk model, and the constraints on the MBH spin and
consequently the radiative efficiency may be also simultaneously
obtained. In the future, if observations can obtain the spectra for a
large sample of QSOs, with good wavelength, luminosity and redshift
coverage, it may be possible to estimate their radiative efficiencies
in a more self-consistent way and then further investigate the
relationship, if any, between the radiative efficiency and the MBH
mass, which would shed light on not only the assembly history of MBHs
\citep[e.g.,][]{Dubois13} but also the physical reasons for the small
spread in the SED shapes of QSOs pointed by \citet{LD11}.

\vskip 0.5cm
{\large \bf \noindent Acknowledgments}
\vskip 0.25cm

\noindent We thank Qingjuan Yu for constructive suggestions and
discussions, Andreas Schulze and Changshuo Yan for helpful
discussions. This work was supported in part by the National Natural
Science Foundation of China under nos. 10973017 and 11033001, the
National Basic Research Program (973 Programme) of China (Grant
2009CB824800), and the BaiRen program from the National Astronomical
Observatory of China, Chinese Academy of Sciences.

\end{document}